\documentclass[useAMS]{mn2e}
\usepackage{times}

\input{epsf}

\title[Observing the effects of the event horizon in black holes]
{Observing the effects of the event horizon in black holes}

\author[C. Done, M. Gierli\'nski] 
{Chris Done$^1$ and Marek~Gierli\'nski$^{1,2}$\\
$^1$Department of Physics, University of Durham, South Road, Durham DH1 3LE, 
UK\\ 
$^2$Obserwatorium Astronomiczne Uniwersytetu Jagiello{\'n}skiego, 30-244 
Krak{\'o}w, Orla 171, Poland}

\date{Submitted to MNRAS}
\pagerange{\pageref{firstpage}--\pageref{lastpage}}
\pubyear{2002}

\begin{document}

\topmargin = -0.5cm

\maketitle

\label{firstpage}

\begin{abstract}

The key difference between neutron stars and black holes is the
presence/absence of a solid surface.  Recent attempts to detect this
difference have concentrated on the quiescent luminosity, but here
these sources are {\em faint} and difficult to observe. Instead we
look at these sources when they are {\em bright}, and show that there
is a clear difference between black holes and neutron stars in the
evolution of their X-ray spectra which is due to the presence of a
surface in the neutron stars. We also show that there is a type of
X-ray spectrum which is {\em only} seen from black holes, making it a
good diagnostic for the nature of new transient sources.

\end{abstract}

\begin{keywords}
  accretion, accretion discs -- X-rays: binaries
\end{keywords}

\section{Introduction}

Black holes and neutron stars have very similar gravitational 
potentials as neutron star radii are of the order of three 
Schwarzchild radii, the last stable orbit of material around a 
black hole. Thus the accretion flows should be similar, except 
that neutron stars have a solid surface, while black holes do 
not. This fundamental difference should give rise to some 
observable effects. Observations of X-ray bursts (from nuclear 
burning of the accreted material onto the surface), or coherent 
pulsations (from a magnetic field) are unique signatures of 
neutron stars.  However, not all neutron star systems show these 
-- they are a sufficient but not necessary condition of a 
surface (e.g.\ the review by van der Klis 1995).

Another key to the presence of a surface is that a boundary 
layer between the accretion flow and the surface can form. In 
Newtonian gravity an accretion disc can radiate only half of the 
gravitational potential energy. The other half is stored as 
kinetic energy of the rotating material, but this must be 
radiated at the surface in a boundary layer if the surface is 
stationary. In general relativity the energy in the boundary 
layer is even larger, about twice that of the disc (Sunyaev \& 
Shakura 1986; Sibgatullin \& Sunyaev 2000). Neutron stars can of 
course be rapidly rotating, but even the fastest spinning 
millisecond pulsar (at 640 Hz) is rotating at approximately half 
the Keplarian period (Backer et al.~1982), where the energy 
released in the boundary layer should be as much as that in the 
disc (Sibgatullin \& Sunyaev 2000).

Attempts to look for the effects of the surface on the spectrum 
have concentrated on comparing black holes and neutron stars in 
quiescence.  At low mass accretion rates the accreting material 
is not very dense, so can form a two temperature plasma (protons 
much hotter than the electrons) rather than the single 
temperature assumed by the standard Shakura-Sunyaev disc models. 
The two temperature accretion flows can cool by Compton 
scattering (Shapiro, Lightman \& Eardley 1976), and by advection 
(carrying the energy along with the flow; Narayan \& Yi 1995). 
The flow is radiatively inefficient if advective cooling 
dominates so the ratio of boundary layer to accretion flow 
luminosity should be very high.  This was the motivation for the 
quiescent luminosity comparisons of black holes and neutron 
stars. While the neutron stars are somewhat more luminous than 
the black holes systems (Narayan, Garcia \& McClintock 1997; 
Garcia et al.~2001), the difference is nothing like as large as 
predicted by the ADAF models (Menou et al.~1999). The 
discrepancy is probably due to the fact that convection and 
outflows are also important, and change the structure (but not 
the existence) of the hot flow (Hawley \& Balbus 2002).

The boundary layer should be very obvious in the neutron star 
spectra at high mass accretion rates, where the sources are {\em 
bright}, and the only known solution of the accretion flow 
equations is a disc. Indeed, the neutron star spectra are 
generally interpreted in terms of disc and boundary layer 
emission, although the spectral decomposition is difficult 
(e.g.\ Mitsuda et al.~1984, 1989; White et al.~1988 but see di 
Salvo et al.~2000a and Done, {\.Z}ycki \& Smith 2002).  However, 
a {\it unique\/} black hole spectral signature remained elusive. 
Proposed spectral characteristics of black holes included the 
ultrasoft spectral state seen at high luminosities (White \& 
Marshall 1984; White et al.~1984), the hard X-ray emission seen 
at low luminosities (e.g. Sunyaev et al.~1991), and the steep 
X-ray tail seen at high luminosities (e.g. Laurent \& Titarchuk 
1999). However, counter-examples to all these are known: Cir X-1 
is a neutron star which shows an ultrasoft spectrum (e.g. Tanaka 
\& Lewin 1995), low luminosity neutron stars show hard X-ray 
power law spectra (Barret \& Vedrenne 1994) and high luminosity 
neutron stars show steep X-ray tails (e.g. di Salvo et al.~2001 
and references therein). 

Only subtle spectral differences between the neutron stars and black
holes seemed apparent.  Barret, McClintock \& Grindlay (1996) were
able to show that there is a difference in hard X-ray (20--200 keV)
emission between the luminous neutron stars and black holes, but the
relative insensitivity of high energy detectors again makes these
observations difficult.  By contrast, these does seem to be unique
black hole/neutron star discriminants in the timing signatures (broad
band power spectra: Sunyaev \& Revnivtsev 2000; quasi-periodic
oscillation frequencies: Belloni, Psaltis \& van der Klis 2002), but
the aim of this paper is to look instead at the spectrum, where the
signature of the surface/boundary layer should be {\em obvious} in
neutron stars but {\em absent} in black holes.

In this paper we collate a large sample of high quality data 
from the {\it Rossi X-ray Timing Explorer\/} Proportional 
Counter Array ({\it RXTE}/PCA) observations of black holes and 
low magnetic field neutron stars (atolls and Z sources), and 
show that the source evolution as a function of luminosity is 
very different depending on the nature of the compact object. We 
also show that there is a type of spectrum which is {\em only\/} 
seen in the black hole systems, where the thermal emission from 
a low temperature disc dominates, but with a steep power-law 
tail to higher energies. We argue that this is a unique black 
hole signature, a sufficient (though not necessary) condition 
for any new transient to be identified as a black hole.

We model the source evolution both qualitatively and 
quantitatively with a similar accretion flow in {\em both} black 
holes and neutron stars.  The observed differences in colour 
evolution can be explained by the additional emission from a 
boundary layer/surface in the neutron stars, while the lack of 
this component in the black hole systems implies the presence of 
an event horizon. 

\section{The Data}

While {\it XMM-Newton\/} and {\it Chandra\/} have opened up new 
windows in high resolution X-ray imaging and spectroscopy, {\it 
RXTE\/} gives an unprecedented volume of data on X-ray binary 
systems. To get a broad idea of the range of spectral shapes 
seen quantity as well as quality is important. But this also 
means that analysing individual spectra is very time consuming. 
Colour-colour and colour-intensity diagrams have long been used 
in neutron star X-ray binaries to get an overview of source 
behaviour (e.g.\ Hasinger \& van der Klis 1989).  The problem is 
that colours are often defined using the counts detected within 
a certain energy range, and so depend on both the instrument 
response and on the absorbing column towards the source (see 
e.g.\ Kuulkers et al.~1994). To get a measure of the true source 
behaviour we want to plot {\em intrinsic\/} colour, i.e.\ 
unabsorbed {\em flux\/} (as opposed to counts) ratios over a 
given energy band.  To do this we need a physical model. 
Plainly, we expect emission from an accretion disc, together 
with a higher energy component from Comptonization. Reflection 
of this emission from the surface of the accretion disc can also 
contribute to the spectrum (see e.g. Done et al.~1992; Yoshida 
et al.~1993; Done, {\.Z}ycki \& Smith 2002 for observations of 
reflection in black holes, neutron star atoll and Z sources, 
respectively). Thus we use a model consisting of a multicolour 
accretion disc, Comptonized emission (which is {\em not\/} a 
power-law at energies close to either the seed photon 
temperature or the mean electron energy), with Gaussian line and 
smeared edge to roughly model the reflected spectral features, 
and Galactic absorption.

The {\it RXTE\/} archival data for a given source is stored in 
segments, generally corresponding to between one and three 
orbits, and consisting of a few kiloseconds of data. For black 
hole candidates and atolls we extract a single PCA spectrum for 
each segment. Due to fast variability of Z sources, shorter than 
typical {\it RXTE\/} orbital timescales, we extract their 
spectra from 512 s intervals of the data. We use {\sc ftools} 
ver.\ 5.2 to extract spectra and background data for PCA 
detectors 0, 2 and 3, top layer only, and add 1 per cent 
systematic error onto each energy channel. We fit these spectra 
to the model described above.  We choose 4 energy bands, 3--4, 
4--6.4, 6.4--9.7 and 9.7--16 keV, and integrate the {\em 
unabsorbed\/} model fluxes over these ranges to form {\em 
intrinsic\/} colours. We define hard colour as a ratio of fluxes 
in 9.7--16 and 6.4--9.7 keV bands and a soft colour as a ratio 
of 4--6.4 and 3--4 keV. We estimate errors on colours assuming 
that the relative error on the intrinsic colour is equal to an 
error of count rate ratio in the same energy ranges. We exclude 
from the analysis the data with poor statistics, where the 
errors on colours exceed 0.1. 

This technique gives results which are, to some extent, model 
dependent. However, as long as a model fits the data well, the 
resulting colours are robust. And indeed, the model described 
above fits all the data with $\chi^2/\nu < 1.5$ (for majority of 
observations $\chi^2/\nu < 1$). To assess the model-dependence 
of the colours we have fit all the data with a simpler model 
consisting of a single-temperature blackbody, power-law and a 
Gaussian line. This model does not always give an adequate fit, 
however where it does ($\chi^2/\nu < 1.5$) the differences in 
derived colours are always smaller than 0.1 and are generally 
smaller than the statistical errors. We have also tested a more 
complex model, using a reflected spectrum rather than the 
phenomenological line/edge description. Since this is too time 
consuming to fit to all the data we chose individual spectra at 
various points on the colour-colour diagram. The difference in 
derived colours between the refection fits and the 
phenomenological line/edge model are smaller than the 
statistical errors. Even taking the reflected emission away from 
the model, i.e. using the true continuum shape, gives colours 
which are $\la 0.1$ different from the Gaussian/smeared edge 
colours. Hence the colours obtained with this method are robust, 
and do not depend on the detailed model used for the line and 
edge features.

Another possible source of inaccuracy can be the interstellar 
absorption, which affects measurement of the soft colour. We 
calculate the colours from unabsorbed model spectrum, but if the 
absorption used in the model is incorrect then this can shift 
the whole diagram along the soft colour axis. A change in 
absorption by $1 \times 10^{22}$ cm$^{-2}$ changes the 
calculated soft colour by $\sim$ 0.1. This is particularly 
important for highly absorbed sources. To avoid these problems 
we only use sources where the absorption is $\le 2\times 
10^{22}$ cm$^{-2}$.

\begin{figure}
\begin{center}
\leavevmode
%{\psfig{file=colcol_baselines.eps,width=5.8cm}}
\epsfxsize=5.8cm \epsfbox{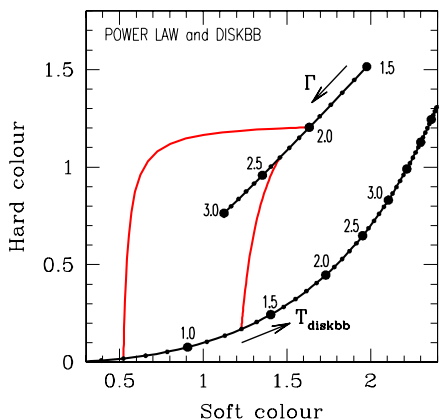}
\end{center}

\caption{The intrinsic colours expected from a power-law 
spectrum with photon spectral index $\Gamma$ = 1.5--3, and that 
formed by a disc blackbody with temperature at the inner disc 
radius $kT_{\rm disc}$ = 0.6--4 keV. The power-law forms a well 
defined diagonal track, while the disc blackbody track curves 
upwards in hard and soft colour as the temperature increases. We 
also plot the track for two composite (disc plus power-law) 
models. The left hand track has a disc with $kT$ = 0.6 keV with 
an increasing fraction of the luminosity (from 0 to 1) in a 
power-law of index $\Gamma$ = 2, while the right hand track is 
for a disc temperature of 1.3 keV and $\Gamma$ = 2.3.}

\label{fig:colcol_baselines}
\end{figure}

For reference we show in Fig. \ref{fig:colcol_baselines} the 
intrinsic colours expected from a power-law and disc blackbody 
spectra. These curves are overlayed in all the figures with 
colour-colour diagrams later in this paper.

We estimate the bolometric flux by extrapolating the unabsorbed 
model, and integrating over all energies.  Obviously this is 
much more model dependent than the technique used to derive the 
colours. We fix a lower limit to the disc temperature of $\ge 
0.5$ keV and an upper limit to the Comptonizing electron 
temperature of $\le 100$ keV so that the continuum components 
cannot produce arbitrarily large luminosities peaking outside of 
the observed bandpass. With these constraints the total model 
flux can be always recovered by integrating between 0.01--1000 
keV. A more restricted band of 0.1--100 keV is adequate for the 
soft states, but can underestimate the low/hard state luminosity 
by up to 40 per cent as the spectrum peaks at 100--200 keV 
(e.g.\ Gierli{\'n}ski et al.~1997).  While this is outside the 
{\it RXTE\/} PCA bandpass, we stress that we use proper 
Comptonization models which roll over at energies above electron 
temperature. The upper limit on the electron temperature of 100 
keV means that the model flux converges with increasing the 
upper integration limit, unlike a power law.

The total model flux, together with the distance to the source, 
$D$, gives an estimate of the bolometric luminosity. Using the 
(fairly well constrained) mass, $M$, we can then translate the 
bolometric luminosity into a fraction of the Eddington 
luminosity, $L/L_{\rm Edd}$, where $L_{\rm Edd} = 1.26 \times 
10^{38} (M / $M$_\odot)$ erg s$^{-1}$.

\section{The Sample}

The aim of this paper is to look at the accretion flow as a 
function of $L/L_{\rm Edd}$ in both neutron stars and black 
holes.  To cover a large range in $L/L_{\rm Edd}$, and hence to 
include all the spectral states, we could either use a large 
sample of objects which vary by only a small amount, or a 
smaller sample of objects which vary by a large amount (or 
both). The key advantage of using single, variable, sources is 
that observed changes in spectral state {\em must} be associated 
with changes in the  accretion flow, rather than with any other 
source property. Hence our main selection criteria is to use 
sources which vary substantially so that they cover a range in 
spectral states. We also require that the absorption is less 
than $2\times 10^{22}$ cm$^{-2}$ so that the intrinsic colours 
are robust.  This latter condition is actually quite a stringent 
constraint since many of the potential sources are at a large 
distance in the Galactic plane.

For the neutron stars, the only low-magnetic field systems 
(atolls) with {\it RXTE\/} data which have $N_H < 2\times 
10^{22}$ cm$^{-2}$ and which sample all three branches of the 
island to banana pattern are Aql X-1, 4U~1608--52 and 
4U~1705--44 (see Muno, Remillard \& Chakrabarty 2002; 
Gierli{\'n}ski \& Done 2002). We also add SAX~J1808.4--3658 
which varied by a factor of 10 in luminosity, but showed no 
strong spectral variability. Neutron stars in low-mass X-ray 
binaries are probably all of similar mass and spin ($1.4$ 
M$_\odot$ and $a_*\sim 0.4$ see e.g.\ the review by van der Klis 
2000), so their spectra should only depend on the (time 
averaged) $L/L_{\rm Edd}$.

We also include the low absorption Z sources, and the peculiar 
neutron star system Cir X-1 to show the range of neutron star 
behaviour.

For the black holes there are rather more objects which show 
state transitions, due to black holes being preferentially 
transient systems (King, Kolb \& Szuszkiewicz 1997). Those which 
also have low absorption and are well observed are Cyg X-1, LMC 
X-3, GX 339--4, XTE~J1550--564, XTE~J1859+226 and GRO~J1655--40. 
%There is evidence for high spin in GRO~J1655--40 (Zhang, Cui \& 
%Chen 1997; Gierli{\'n}ski, Macio{\l}ek-Nied{\'z}wiecki \& 
%Ebisawa 2001), whereas the other black holes used here are 
%probably only moderately rotating (Cui, Zhang \& Chen 1998), 
%similarly to the neutron stars. For a direct comparison of 
%accretion flow properties we want to be able to match the black 
%hole and neutron star samples as closely as possible, so we 
%exclude GRO~J1655--40 from this paper.

Table \ref{tab:sources} shows the systems which we used, 
together with the mass and distance assumed for calculating 
$L/L_{\rm Edd}$. For illustration purposes we also quote some 
uncertainties of mass and distance found in the literature, 
however we don't use them in this paper. Obviously the Eddington 
limits for sources such as XTE~J1550--564, GX 339--4 and Cir X-1 
are poorly constrained, while those like Cyg X-1 and LMC X-3 are 
much more secure. The table also includes the assumed Galactic 
absorption.

We use all the {\it RXTE\/} PCA (detectors 0, 2 and 3) data of 
these sources which is publicly available (as for December 
2002), except for Cir X-1 and GRO~J1655--40. Both of these have 
variable absorption (e.g.\ Shirey, Bradt \& Levine 1999; 
Kuulkers et al.~1998), so for these sources we select the data 
not affected by this from the lightcurves and spectra. For 
GRO~J1655--40, detector 3 of the PCA was often off, so we use 
data from detectors 0 and 2 to follow the outburst behaviour. 

\begin{table*}
\begin{tabular}{lccccc}
\hline 
Source Name & $M$ (M$_\odot$) & $D$ (kpc)
& $N_H$ ($10^{22}$ cm$^{-2}$) & References \\
\hline
Cyg X-1 & 10 (4.8--14.7) & 2 (1.8-2.2) & 0.6 & 1, 2, 3\\
GX 339--4 & 6 (2.5--10) & 4 (2.6--5) & 0.6 & 4, 5 \\
XTE J1550--564 & 10 (9.7--11.6)& 5.3 (2.8--7.6) & 0.4 & 6, 7 \\
LMC X-3 & 9 (7--14)& 52 (51.4--52.6) & 0.06 & 8, 9, 10 \\
XTE J1859+226 & 10 (5--12) & 7.6 (4.8--8) & 0.4 & 11 \\
GRO J1655--40  & 7 (6.8--7.2) & 3.2 (3--3.4) & 0.8 & 12, 13, 14 \\
\hline
4U 1608--522 & 1.4 & 3.6 (2.2--5) & 1.5 & 15, 16, 17 \\
4U 1705--44 & 1.4 & 7 (6.3--8.2) & 1.2 & 18, 19 \\
Aql X-1 & 1.4 & 5 (4--6.5) & 0.5 & 20, 21 \\
SAX J1808.4--3658 & 1.4 & 2.5 (2.2--3.2) & 0.1 & 22, 23 \\
\hline 
Sco X-1 & 1.4 & 2.8 (2.5--3.1) & 0.15 & 24, 25 \\
GX 17+2 & 1.4 & 8 (5.6--10.4) & 2 & 26, 27 \\
Cyg X-2 & 1.4 & 11.6 (11.3--11.9) & 0.22 & 19, 28 \\
\hline  
Cir X-1 & 1.4 & 10 ($ >$ 8) & 2 & 19, 29, 30 \\ 
\hline
\end{tabular}

\caption{The list of the sources used in this paper, together 
with assumed mass ($M$), distance ($D$) and absorption column 
($N_H$) and their uncertainties. The numbered references are as 
follows:
[1] Herrero et al.~1995
[2] Gierli{\'n}ski et al.~1999
[3] Ba{\l}uci{\'n}ska-Church et al.~1995
[4] Cowley et al.~2002
[5] Zdziarski et al.~1998
[6] Orosz et al.~2002
[7] Wilson \& Done 2001
[8] Cowley et al.~1983
[9] di Benedetto 1997
[10] Haardt et al.~2001
[11] Hynes et al.~2002
[12] Orosz \& Bailyn 1997
[13] Hjellming \& Rupen 1995
[14] Gierli{\'n}ski et al. 2001
[15] Nakamura et al.~1989, 
[16] Wachter et al.~2002
[17] Penninx et al.~1989
[18] Haberl \& Titarchuk 1994
[19] Predehl \&  Schmitt 1995
[20] Rutledge et al.~2001
[21] Church \& Ba{\l}uci{\'n}ska-Church 2001
[22] in 't Zand et al.~2001
[23] Campana et al.~2002
[24] Bradshaw, Fomalont \& Geldzahler 1999
[25] Paerels, Kahn \& Wolkovitch 1998
[26] Kuulkers et al.~2002
[27] di Salvo et al.~2000b
[28] Orosz \& Kuulkers 1999
[29] Goss \& Mebold 1977
[30] Mignani et al.~2002
}

\label{tab:sources}
\end{table*}

%################################################################
 
\section{Results}

\subsection{Black holes}

\begin{figure*}
\begin{center}
\leavevmode 
%{\psfig{file=colum_bh.eps,width=10cm}}
\epsfxsize=10cm \epsfbox{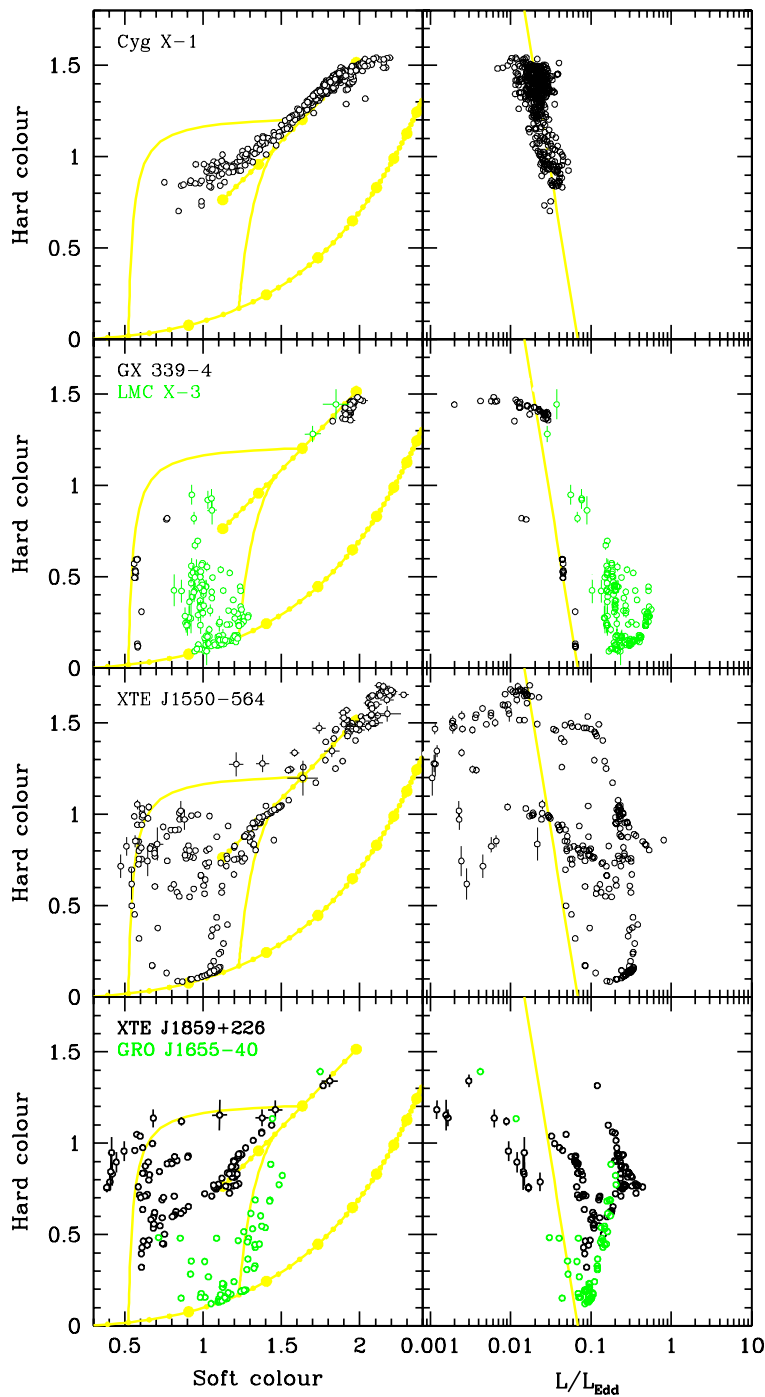}
\end{center}

\caption{Colour-colour (left) and colour-luminosity (right) 
plots for the {\it RXTE}/PCA data on the Galactic black holes 
Cyg X-1 (top panel), GX 339--4 and LMC X-3 (second panel), 
XTE~J1550--564 (third panel) and XTE~J1859+226 and GRO~J1655--40 
(lower panel). The best fitting line to the Cyg X-1 
colour-luminosity plot is reproduced on the colour-luminosity 
plots for the other black holes. The colour-colour diagrams also 
include the tracks indicating the colours produced by a 
power-law and disc blackbody (see 
Fig.~\ref{fig:colcol_baselines}).}

\label{fig:colum_bh}
\end{figure*}

The left hand panel in Fig.~\ref{fig:colum_bh} shows the 
intrinsic colours, while the right hand panel shows the hard 
colour versus the reconstructed $L/L_{\rm Edd}$.  The top panels 
show data from Cyg X-1, where the range in luminosity is small 
and it traces out a well defined track in both the colour-colour 
and colour-luminosity plots as it switches from the low/hard to 
high/soft state (see e.g.\ the review of black hole spectral 
states in Tanaka \& Lewin 1995; Esin, McClintock \& Narayan 
1997). The second panel shows GX 339--4 and LMC X-3 which 
together span a much larger range in colour and intensity. Their 
tracks look very similar to those in the third panel from the 
best observed black hole transient, XTE~J1550--564, and to the
behaviour of GRO~J1655--40 and XTE~J1859+226 in the lower panel.
All these black holes are consistent with the same spectral evolution, 
in which low/hard state spectra form a well defined diagonal track in 
colour-colour plots, while the high/soft spectra show an amazing 
variety of shapes for a given $L/L_{\rm Edd}$.

\begin{figure*}
\begin{center}
\leavevmode
%{\psfig{file=colspec_bh.eps,width=13cm}}
\epsfxsize=13cm \epsfbox{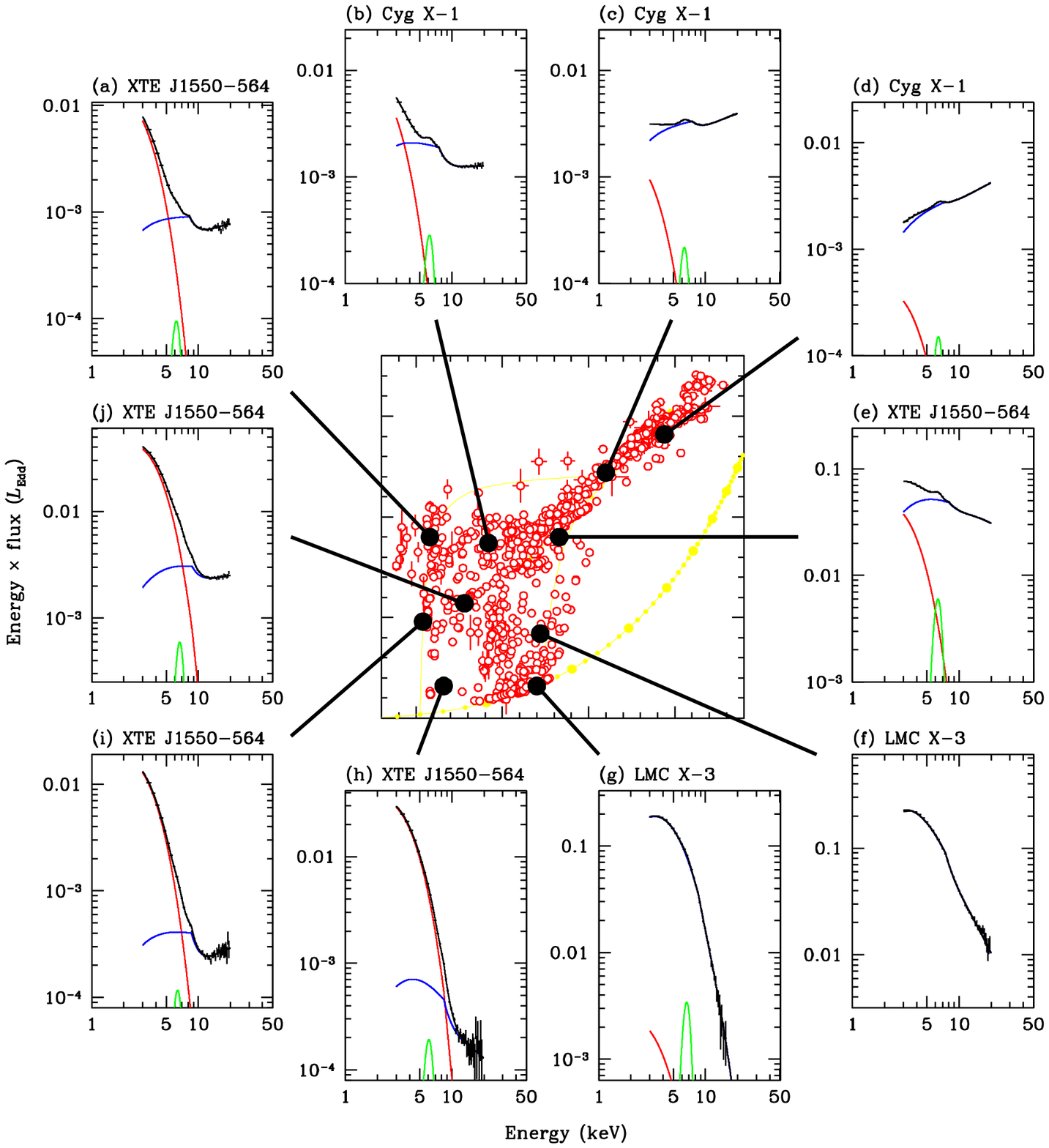}
\end{center}

\caption{In the middle: the colour-colour diagram of the black hole 
candidates, containing all the data from the left-hand panels in 
Fig.~\ref{fig:colum_bh} (with identical axes, for which the labels 
were removed here for the sake of clarity). Around the diagram there 
are unfolded energy spectra corresponding to different points in the 
diagram. Distances and masses from Table \ref{tab:sources} have been 
applied to convert observed fluxes into $L/L_{\rm Edd}$. The spectra 
show the PCA data and best-fitting model components: the disc (the 
soft component), the Comptonization (the hard component) and the 
Gaussian line (at $\sim 6.4$ keV). See electronic edition for the 
colour version of this figure.}

\label{fig:colspec_bh}
\end{figure*}

The colour-colour tracks of the simple models (power-law and disc)
shown in Fig. \ref{fig:colcol_baselines} are overlaid on the data and
show that the well defined diagonal track of the low/hard state
spectra fall on the power-law track. The variety of high/soft state
spectral colours are mostly bounded by the two composite models, and the
disc blackbody track (implying a disc temperature between 0.6--1.3
keV, $\Gamma$ = 2.1--2.3 and a power-law fraction between 0 and 1). The
spectral shape at a range of points on this colour-colour diagram is
shown explicitly in Fig.~\ref{fig:colspec_bh}. For reference, the 
nomenclature used in this paper is that hard 
power-law spectra such as $c$ and $d$ are termed low/hard state,
completely disc dominated spectra ($g$ and $h$) are termed ultrasoft,
disc dominated spectra with some steep power law tail ($a$ and $b$)
are termed high/soft state, while those in which the steep power law
and disc have comparable luminosities (e.g. spectrum $e$) are very
high state.

The colour-luminosity track seen in Cyg X-1 is extrapolated and
overlaid on all the other colour-luminosity plots in
Fig. \ref{fig:colum_bh}.  There are clear indications that all of the
sources can change from the hard to soft state along this track,
although plainly the situation is more complex than a simple switch at
$L/L_{\rm Edd}\sim 0.03$.  XTE~J1550--564 can be in the hard state at
$\sim 0.1 L/L_{\rm Edd}$, {\em and} can be in the soft state at $\sim
0.01 L/L_{\rm Edd}$.  The accretion flow clearly does {\em not} simply
depend on $L/L_{\rm Edd}$ (Nowak 1995; Maccarone \& Coppi 2002b see
also van der Klis 2001 for a similar effect in the neutron star
systems).  

\subsection{Neutron stars: Atolls and Z sources}

\begin{figure*}
\begin{center}
\leavevmode
%{\psfig{file=colum_ns.eps,width=10cm}}
\epsfxsize=10cm \epsfbox{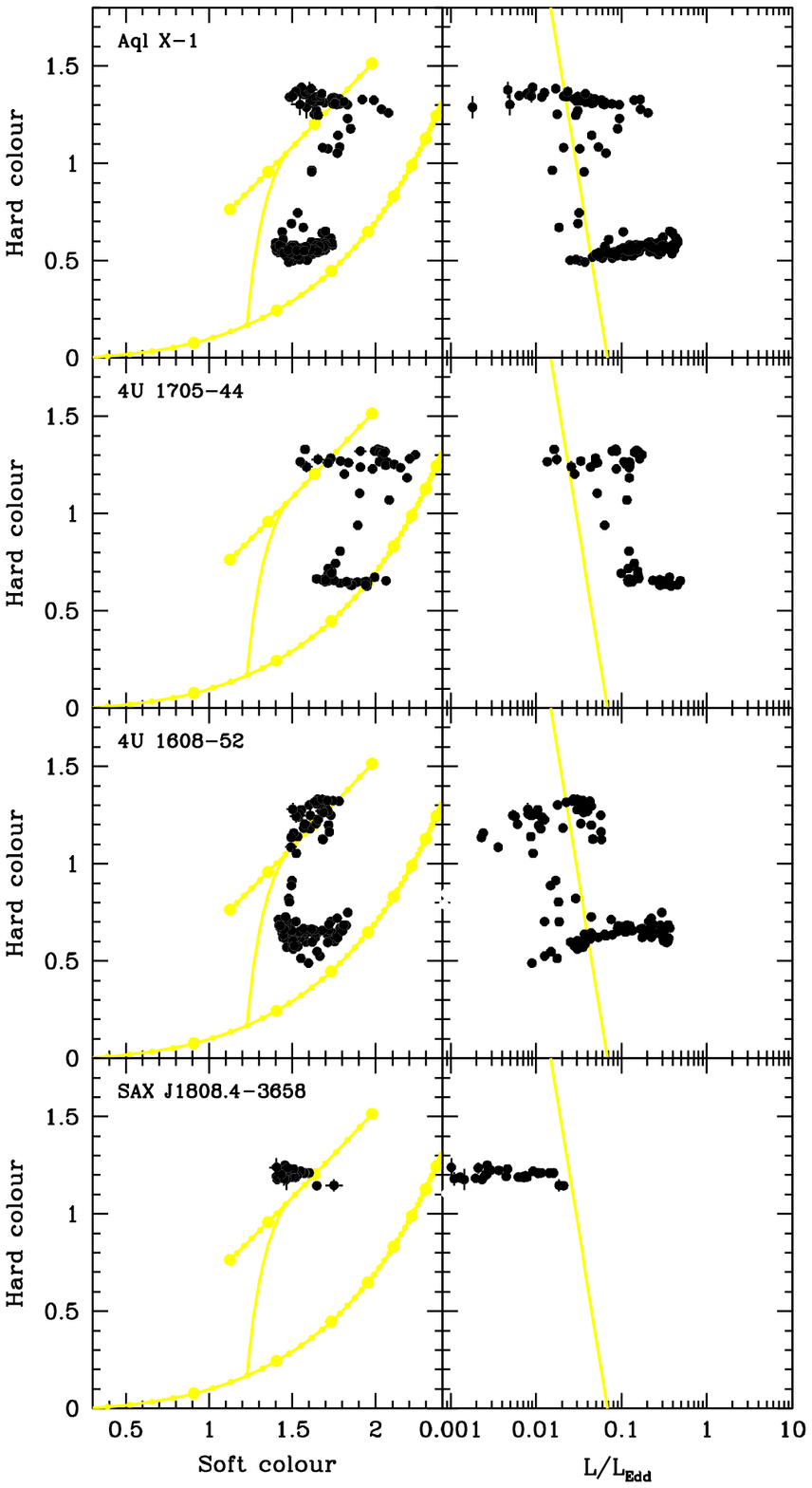}
\end{center}

\caption{Colour-colour (left panels) and colour-luminosity 
(right panels) plots for the {\it RXTE}/PCA data on the 
transient atolls 4U~1705--44, 4U~1608--52, Aql~X-1 and 
SAX~J1808.4--3658. The atolls also show a well defined hard-soft 
transition, but the colours and how they evolve as a function of 
$L/L_{\rm Edd}$ are completely different to those from the black 
holes. The disc and power-law model lines for the black holes 
are overlaid onto the colour-colour diagrams for the neutron 
stars for comparison. The hard state for the atolls forms a 
track which moves horizontally to the right with increasing 
$L/L_{\rm Edd}$ while the black hole hard state track moves 
diagonally down and to the left. The best fit line to the Cyg 
X-1 colour-luminosity plot is reproduced on the 
colour-luminosity plots for these atoll systems showing that the 
hard state may persist to somewhat higher luminosities than the 
black holes.}

\label{fig:colum_ns}
\end{figure*}

\begin{figure*}
\begin{center}
\leavevmode
%{\psfig{file=colspec_ns.eps,width=10.1cm}}
\epsfxsize=10.1cm \epsfbox{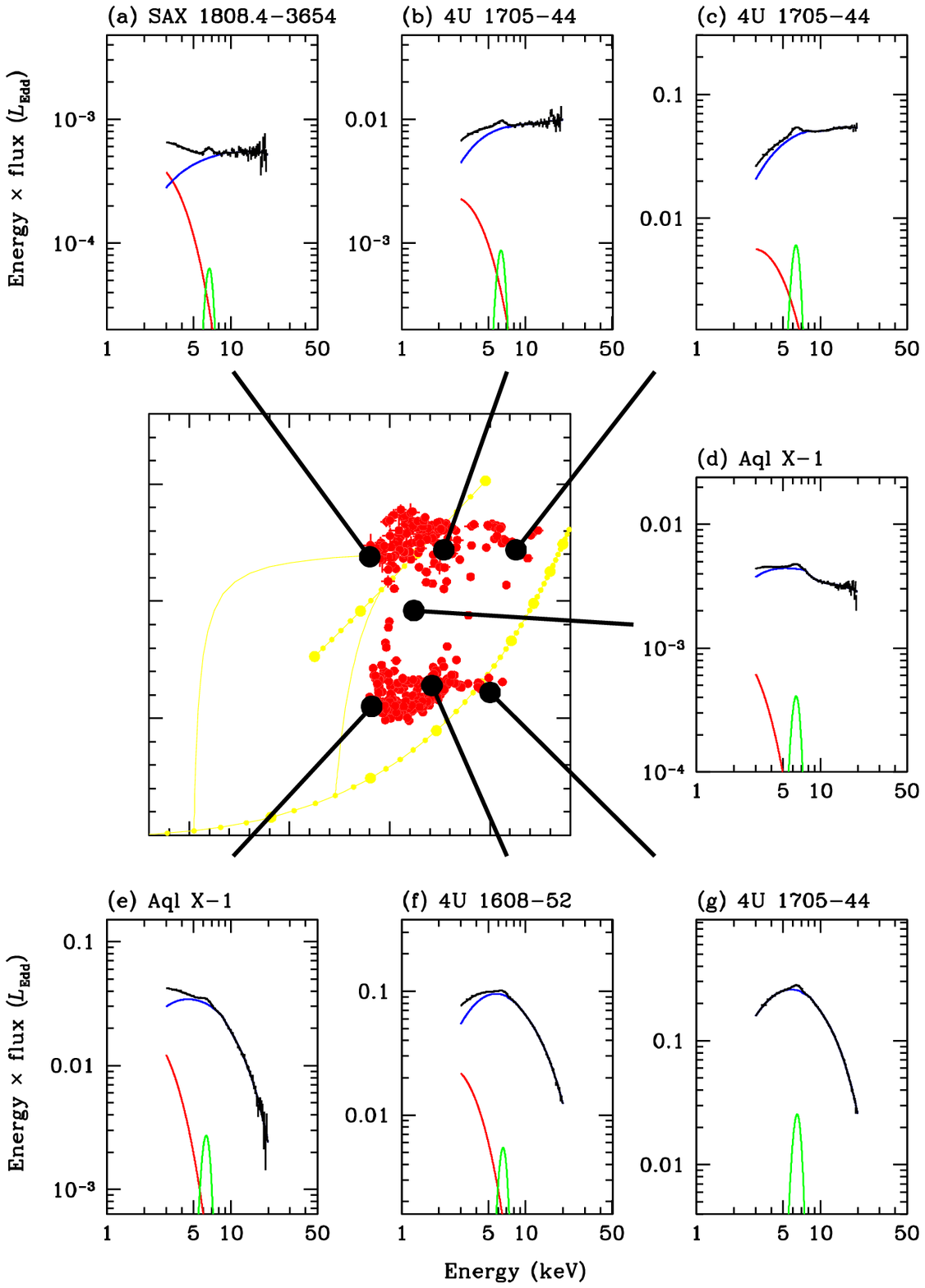}
\end{center}

\caption{As in Fig.~\ref{fig:colspec_bh} but for the 
low-magnetic neutron stars (atolls), containing all the data 
from the left-hand panels in Fig.~\ref{fig:colum_ns}. See 
electronic edition for the colour version of this figure. }

\label{fig:colspec_ns}
\end{figure*}

Fig.~\ref{fig:colum_ns} shows the same colour-colour and
colour-luminosity plots for the atoll neutron star systems, while
Fig. \ref{fig:colspec_ns} shows representative spectra at different
points along the colour-colour track. The intrinsic colour-colour
diagrams are similar for each atoll, with the track forming a large Z
with increasing $L/L_{\rm Edd}$ (Gierli{\'n}ski \& Done 2002a; Muno,
Remillard \& Chakrabarty 2002). This traces a transition from a well
defined hard (island, spectra $a$--$d$ in Fig.~\ref{fig:colspec_ns})
state to a softer spectrum (banana state, spectra $e$--$g$ in
Fig.~\ref{fig:colspec_ns}).  The tracks shown in
Fig. \ref{fig:colcol_baselines} which bounded the black hole colours
are shown overlaid on the colour-colour diagrams for the atoll
systems. This shows that the hard spectra seen in the atolls at low
luminosities (island state, spectra $a$--$d$) have similar colours to
those of the brightest low states seen in the black holes.  This
spectral similarity noted by e.g.\ Yoshida et al.~(1993), Barret et
al.~(1996) and Barret et al.~(2000), contrasts with the very different
spectral evolution of the neutron stars and black hole systems. In the
hard state the atolls move horizontally from left to right with
increasing average flux, while the black holes move diagonally down
and to the left.  After the hard-soft transition the atolls have a
well defined soft track, whereas the black holes show a variety of
spectral states.

\begin{figure*}
\begin{center}
\leavevmode
%{\psfig{file=colum_z.eps,width=10cm}}
\epsfxsize=10cm \epsfbox{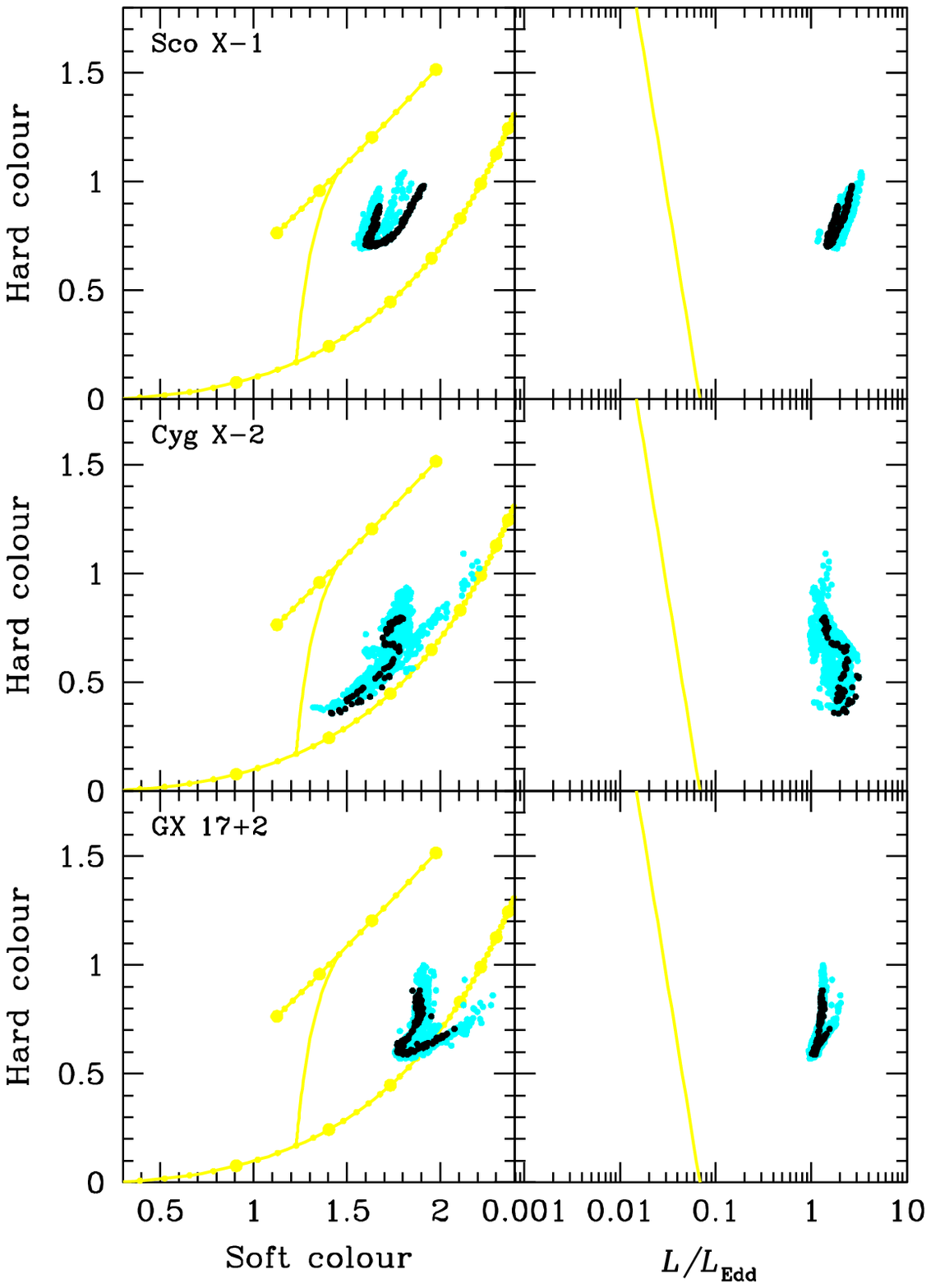}
\end{center}

\caption{Colour-colour (left panels) and colour-luminosity 
(right panels) plots for the {\it RXTE}/PCA data on the Z 
sources Sco X-1, Cyg X-2 and GX 17+2. The Z sources cover rather 
limited space both in colours and luminosity, when compared to 
atolls and black holes. Due to a secular motion in the diagrams 
the 'Z' pattern is blurred, so we plot a selection of data in 
black, to show the pattern traced out by these source in a short 
period.}

\label{fig:colum_z}
\end{figure*}

Fig.~\ref{fig:colum_z} shows the same information for the Z 
sources. By contrast to the very variable black holes and 
transient atolls used in this paper, the Z sources are stable to 
the disc instability because of their high mass transfer rate 
(King \& Ritter 1998). Their typical luminosities are around the 
Eddington luminosity, and they vary only by factors of a few, 
rather than by several orders of magnitude as seen in the 
transient systems. The Z pattern is not particularly evident in 
any of the individual sources, due to the secular shift in the 
pattern over time. These shifts are poorly understood but may be 
due to obscuration (Kuulkers et al.~1994; Kuulkers, van der Klis 
\& Vaughan 1996) or changes in the overall mass accretion rate 
(Homan et al.~2002).  We plot a section of contiguous data for 
each Z source which shows the track more clearly.

In contrast to the black hole and atoll systems, the Z sources 
do not overlap with each other on the colour-colour diagram. In 
addition, the secular shift in the position of the track smears 
out the track for each source. Differences in magnetic field 
and/or spin period may be the most obvious way to explain this, 
since Sco X-1 and GX~17+2 are thought to be at similar 
inclinations (Kuulkers et al.~1996). 

\begin{figure*}
\begin{center}
\leavevmode
%{\psfig{file=colum_cir.eps,width=10cm}}
\epsfxsize=10cm \epsfbox{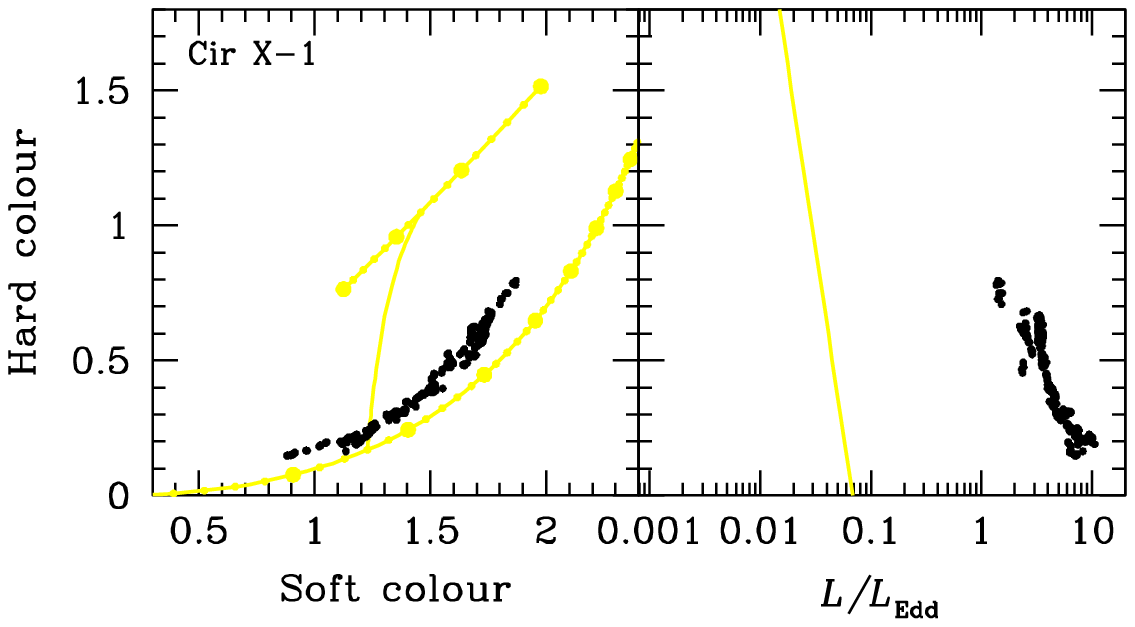}
\end{center}

\caption{Colour-colour (left panels) and colour-luminosity 
(right panels) plots for the {\it RXTE}/PCA data on Cir X-1. The data 
have been selected excluding periods of increased absorption.}

\label{fig:colum_cir}
\end{figure*}

Fig~\ref{fig:colum_cir} shows the colour-colour and 
colour-luminosity data for Cir X-1. This system has shown X-ray 
bursts so must be a neutron star (Tennant, Fabian \& Shafer 
1986), and is tentatively identified as a Z source (Shirey et 
al.~ 1999). There is only a lower limit to its distance ($\ge 8$ 
kpc: Goss \& Mebold 1977) as recent {\it HST\/} observations 
have shown that it is not associated with a nearby supernova 
remnant (Mignani et al.~2002). It is highly supereddington, with 
luminosity of $L/L_{\rm Edd}$ = 2.5--10 for a 1.4 solar mass 
neutron star at 10 kpc. With these parameters its colour-colour 
and colour-luminosity diagrams are similar to parts of those of 
Cyg X-2, but extend to much higher luminosities, where it shows 
spectra unlike those of any other neutron star (Z or atoll). 
This seems to give a consistent picture, in which Cir X--1 is a 
Z source (i.e. a neutron star with moderate magnetic field) but 
is seen at generally higher luminosities than the `normal' Z 
sources, so can have quite different spectral evolution. 

\begin{figure}
\begin{center}
\leavevmode
%{\psfig{file=colcol_all.eps,width=7cm}}
\epsfxsize=7cm \epsfbox{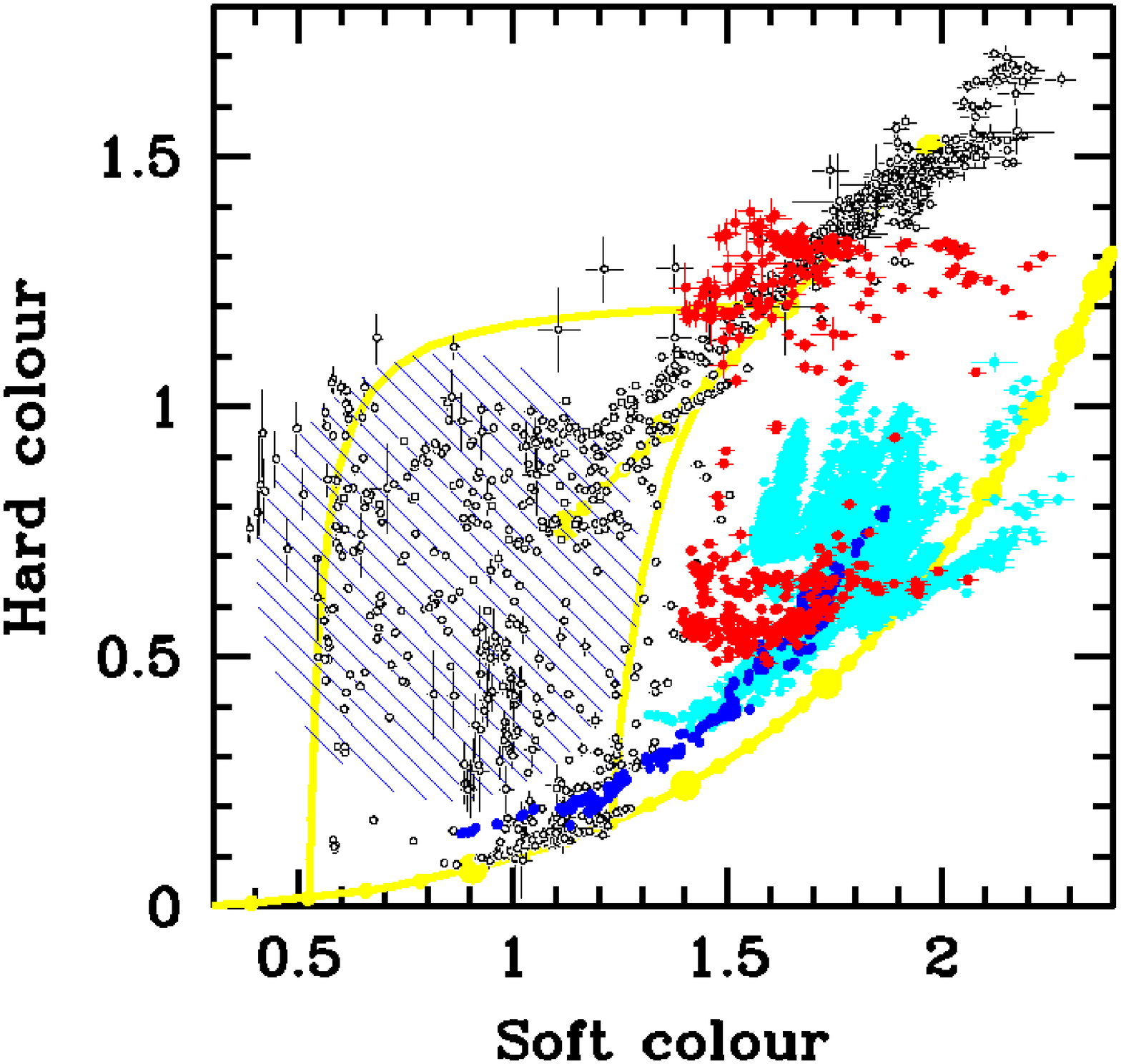}
\end{center}

\caption{Combined colour-colour diagrams for all source types: 
open circles for black holes, red, cyan and blue filled circles 
for atolls, Z sources and Cir X-1, neutron star systems, 
respectively. Hatched region is {\em inaccessible\/} to neutron 
stars due to their boundary layer emission. Spectra with colours 
falling in this region are seen {\em only\/} in black holes.}

\label{fig:colcol_all}
\end{figure}

Fig.~\ref{fig:colcol_all} shows all the data for black holes, 
atolls and Z sources overlaid. It is clear that the various 
classes of sources have regions of overlap. The high luminosity 
atolls (banana branch) show similar spectra to the Z sources (a 
similarity noted many times before e.g.\ van der Klis 1995). 
Low-luminosity black holes and atolls can both show hard colours 
$\ge 1$ while black holes at the highest luminosity show similar 
colours to atolls on the island/banana branch transition. More 
importantly, at highest luminosity Cir X-1 ends up at the same 
place on the colour-colour diagrams as the ultrasoft black hole 
spectra. It is clear that there are {\em some} parts of the 
colour-colour diagram where both neutron stars and black holes 
can be found. 

However, {\em no\/} neutron star systems show any spectrum which 
has soft colour between 0.5--1.1 and hard colour $>0.5$ (hatched 
region in Fig.~\ref{fig:colcol_all}). These types of spectra, 
with small soft colour from a dominant low temperature disc 
spectrum, and mid hard colour from the power-law tail (high/soft 
state) seem to be a unique black hole signature.  The steep 
power law tail in these high/soft state spectra was previously 
suggested as a black hole identifier (e.g. Laurent \& Titarchuk 
1999), but recent observations show such components in the Z 
sources (see, for example, di Salvo et al.~2001 and references 
therein). Very soft spectra (low soft colour) were proposed as a 
unique black hole signature (White \& Marshall 1984), but this 
criterium also includes the ultrasoft spectra which can be 
produced by the neutron star Cir X-1. 

Here we propose that the high/soft state spectra are a 
sufficient (although not necessary) black hole signature, and 
that this region of the colour-colour diagram (see 
Fig.~\ref{fig:colcol_all}) is inaccessible to the neutron star 
systems simply because they have additional boundary layer 
emission.  Even if the accretion flow around a neutron star gave 
rise to an identical spectrum as that around a black hole, with 
a low temperature disc and steep X-ray tail (though the disc 
would be at a marginally higher temperature due to the smaller 
mass of the neutron star), the neutron star should have 
additional emission from the boundary layer. At the high mass 
accretion rates above the hard-soft transition the boundary 
layer should be optically thick, and mostly thermalized (Popham 
\& Sunyaev 2001). This will give an additional higher 
temperature component in the neutron star spectra, increasing 
its soft colour out of the range seen from the black hole 
systems.

To summarise, there are observational differences between the 
black holes and neutron star systems. They evolve very 
differently as a function of average flux on a colour-colour 
diagram, and black holes can occupy regions on this diagram 
which no neutron star can get to. The obvious interpretation is 
that there are physical, observable spectral differences due to 
the presence/absence of a solid surface.

%##############################################################

\section{Simple source evolution models}

Specific models of a truncated disc/inner hot flow have been 
used to explain the spectra and spectral transitions in the 
black hole binaries (Shapiro et al.~1976; Esin et al.~1997). 
These models can also qualitatively explain the shape of the 
variability power spectrum (Churazov, Gilfanov \& Revnivtsev 
2001), and the correlations between the characteristic power 
spectral break frequencies and QPO's (Psaltis \& Norman 2000; 
Stella \& Vietri 1998). Truncation of the disc is also required 
for the disc instability model for the black holes to produce 
long quiescent periods, rather than many mini-outbursts (Dubus, 
Hameury \& Lasota 2001) and is compatible with all the 
constraints on the extent of the accretion disc as measured by 
direct emission (McClintock et al.~2001) and reflection 
({\.Z}ycki, Done \& Smith 1998, Gilfanov, Churazov \& Revnivtsev 
1999). 

Here we show how a phenomenological truncated disc/inner hot 
flow model can explain the evolution of both black holes and 
neutron star binaries on the colour-colour and colour-luminosity 
diagram. For all types of sources we assume that the evolution 
of the source can be explained if the main parameter driving the 
spectral evolution is the average mass accretion rate, $\dot{m}$ 
(which we define in units of Eddington accretion rate, $\dot{m} 
\equiv \dot{M}c^2 / L_{\rm Edd}$). 

\subsection{Black Hole Binaries}

A qualitative picture could be as follows, starting at low 
$\dot{m}$. The disc truncates at rather large radii, and most of 
the accretion takes place in an optically thin, geometrically 
thick hot inner flow. There are few seed photons for Compton 
scattering so the X-ray spectrum is hard 
(Fig.~\ref{fig:colspec_bh}$d$). As $\dot{m}$ increases, the 
truncation radius of the disc decreases, so it penetrates 
further into the hot flow. The disc is brighter and hotter, and 
the changing geometry gives a higher fraction of soft photons 
which are intercepted by the hot flow. This increases the 
Compton cooling, so the electron temperature is lower and the 
Comptonized spectrum steepens (Fig.~\ref{fig:colspec_bh}$c$). 
Both hard and soft colours decrease together as the X-ray 
spectrum in the PCA bandpass is still dominated by the Compton 
scattered power-law. When the inner flow becomes optically thick 
it collapses, and the inner disc replaces the hot flow (see e.g. 
Esin et al.~1997).  The disc then is hot enough to contribute to 
the spectrum above 3 keV, so the soft colour decreases 
(Fig.~\ref{fig:colspec_bh}$b$). The hard X-ray tail is produced 
by the small fraction of magnetic reconnection which takes place 
outside of the optically thick disc (high state). At even higher 
$\dot{m}$ the disc structure is not well understood, but there 
seems to be a variety of disc states, ranging from ones 
dominated by the disc emission (ultrasoft:
Fig.~\ref{fig:colspec_bh}$g$), to ones in which there is both 
the disc emission and a non-thermal power-law tail extending to 
high energies (high/soft state: Fig.~\ref{fig:colspec_bh}$b$), and 
ones in which the disc emission is Comptonized by a low 
temperature thermal plasma as well as by the non-thermal 
electrons (very high state: Fig.~\ref{fig:colspec_bh}$e$).

Fig.~\ref{fig:colspec_bh}$d$ shows that the low/hard state 
spectra are very close to a power-law shape, and the overall 
change in colour in this state can be mostly described simply by 
increasing the power-law index with increasing $\dot{m}$. The 
only significant exception to this is the low/hard state spectra 
seen from XTE~J1550--564 on its 8 day rise to outburst in 1998. 
The source was first observed with colour close to that of a 
power-law of index $\Gamma$ = 1.5, and then softens by 0.2 in 
hard colour with little change in soft colour, taking it below 
the power-law track. The colours then change together, forming a 
roughly parallel track to the power-law but offset by 0.2 in soft
colour, until the source curves back to the power-law track at 
$\Gamma=2$ as it approaches it maximum (lower panel on 
Fig.~\ref{fig:colum_bh}). This anomalous behaviour can easily be 
modelled by including a small amount of {\em ionized\/} 
reflection. The increased reflectivity of the ionized material 
means it contributes to both soft and hard colour bands, whereas more 
neutral reflection contributes only to the hard bands. 
Such ionized reflection is seen in XTE~J1550--564 in these rise spectra 
(Wilson \& Done 2001), and might be expected as the disc is 
plainly far from steady state.

Conversely, the ultrasoft spectra are very close to the disc blackbody
track, which forms the lower limit to the hard colour from the
galactic black holes. Any of the rest of the soft state colours can be
made with a combination of a given temperature disc blackbody spectrum
and a given index of power-law.  Rough ranges for the expected disc
temperatures for the soft state spectra from a $\sim 10 $M$_\odot$
Schwarzchild black hole are 0.6--1.6 keV, where the upper limit is for
accretion at the Eddington limit (with colour correction of 1.8:
Shimura \& Takahara 1995; Merloni, Fabian \& Ross 2000) while the
lower limit is taken from the observation that in general the soft
spectra are seen at luminosities above $\sim 0.03 L_{Edd}$ (although
there is some spread here since the X-ray luminosity is not uniquely
related to spectral shape: van der Klis 2001; Homan et al.~2001;
Wijnands \& Miller 2002).

\subsection{Atolls}

\begin{figure*}
\begin{center}
\leavevmode
%{\psfig{file=ns_simple.eps,width=10.5cm}}
\epsfxsize=10.5cm \epsfbox{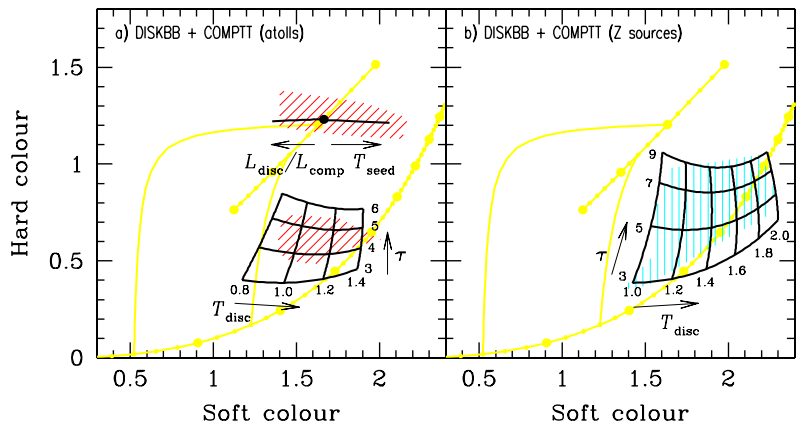}
\end{center}

\caption{Phenomenological source models for the neutron star 
systems evolution on the colour-colour diagram. The shaded areas 
show the range of colours covered by the atolls (panel $a$, see 
also Fig.~\ref{fig:colum_ns}) and Z sources (panel $b$, see also 
Fig.~\ref{fig:colum_z}). The grids in both panels show colours 
predicted by a model consisting of the disc blackbody and 
thermal Comptonization ($kT_e$ = 3 keV) with equal luminosities. 
The disc temperatures shown in the diagram are in units of keV. 
The models for the hard/island state of the atolls (panel $a$, 
upper part) are: power-law ($\Gamma$ = 1.95) with increasing 
contribution from disc blackbody of $kT_{\rm disc}$ = 0.6 keV 
(left of the central dot) and thermal Comptonization ($kT_e$ = 
40 keV, $\tau$ = 1) with increasing temperature of the seed 
photons from 0.3 to 0.85 keV (right of the central dot).}

\label{fig:ns_simple}
\end{figure*}

Unlike black holes, neutron stars have a solid surface so the 
kinetic energy at the innermost disc orbit can also be released 
in a boundary layer.  The luminosity dissipated in the boundary 
layer depends on the accretion flow as well as on $M/R$ and the 
spin of the neutron star. The boundary layer should be 1--2 
times as bright as the disc for a standard disc extending down 
to the neutron star surface (Sunyaev \& Shakura 1986; 
Sibgatullin \& Sunyaev 2000). 

At low $\dot{m}$ the disc is truncated a long way from the 
neutron star, so it emits few soft photons and very few of 
these are intercepted by the inner hot flow. The boundary layer 
is not very dense, so does not thermalize into a blackbody
but instead radiates the energy through Comptonization
(King \& Lasota 1987; Popham \& Sunyaev 2001). 
This hot, optically thin boundary layer 
emission can join smoothly onto the emission from the inner 
accretion flow (Medvedev \& Narayan 2001). The hard X-rays from 
the hot plasma (accretion flow and a boundary layer combined) 
illuminates the neutron star surface. The reflection albedo is 
small at high energies due to Compton down-scattering, so even if 
the neutron star surface is fully ionized it cannot reflect all 
the illuminating flux. The energy absorbed is reprocessed, 
creating a small thermal component which will form the dominant 
source of seed photons for Compton cooling of the inner flow if 
the disc is far away. As $\dot{m}$ increases, the disc moves 
inwards in the same way as for the black holes. However, the 
changing disc geometry has no effect on the fraction of seed 
photons Compton scattered by the hot inner flow if the seed 
photons are still predominantly from the neutron star surface. 
The constant seed photon/hot inner flow geometry irrespectively 
of $\dot{m}$ means the balance of heating and cooling stays the 
same so the Comptonized continuum and hence the high energy 
spectral shape does not change. 

Fig.~\ref{fig:ns_simple}$a$ shows the region of colour-colour 
space occupied by the atolls (diagonal shading), with a solid 
circle showing the colours of a power-law spectrum of $\Gamma$ = 
1.95. Adding any form of soft component from the disc or neutron 
star surface softens the low energy spectrum (reducing the soft 
colour), but does not affect the hard colour, forming a 
horizontal track extending to the left of the power-law track. 
However, it is clear that this cannot explain the large patch of 
atoll island state spectra which are horizontally to the right 
of the power-law track. To match these spectra requires a 
hardening the low energy spectrum, while keeping the power-law 
of $\Gamma$ = 1.95 dominating the higher energy spectrum. From 
direct data fitting it is clear that these large soft colours 
are easily produced when proper Comptonization models are used 
to describe the hard X-ray component, rather than a power-law 
(e.g.\ Gierli{\'n}ski \& Done 2002b). True Comptonization has a 
low energy cutoff close to the seed photon energies (see e.g.\ 
Done et al.~2002) which hardens the soft spectrum. 

We show the track corresponding to a Comptonized 
spectrum with $\Gamma=1.95$ (modelled by {\sc comptt} with 
$\tau=1$ and $kT_e=40$ keV) from seed photons from the neutron 
star surface increasing in temperature from 0.3 to 0.85 keV.
Plainly this matches very well with the island branch to the right of
the power law track. This leads us to explain the 
entire upper island branch can be explained as follows.  At low
$\dot{m}$ the disc is far away from the neutron star so cannot
contribute to the PCA spectrum. The observed soft component, which
pulls the colours to the left of the power-law track is from direct
emission from the neutron star surface which can be seen as the
optical depth of the flow is low.  With increasing $\dot{m}$ the flow
becomes more optically thick and the seed photon temperature from the
neutron star surface increases.  This increasing optical depth reduces
the contribution of the direct soft photons from the neutron star
surface which escape unscattered, so the soft colour
increases. Eventually, none of the seed photons can be seen, but the
increasing seed photon temperature starts to put a low energy cutoff
on the spectrum, hardening the low energy spectrum further. As the
seed photon geometry remains the same throughout all this (dominated
by photons from the neutron star surface rather than from the disc)
then the net result is a horizontal track where the soft colour
increases with source luminosity.

The atolls keep this constant hard colour until the inner 
flow/boundary layer reaches its maximum luminosity when it 
becomes optically thick. This causes the hard colour to decrease 
as the cooling is much more effective when the boundary layer 
thermalizes, so its temperature drops. Since it has more 
luminosity to thermalize over a smaller area, its temperature is 
higher than  that of the disc (Popham \& Sunyaev 2001).  The 
disc replaces the inner hot flow, so the disc temperature starts 
to contribute to the PCA bandpass and the soft colour decreases. 
The track turns abruptly down and to the left during this 
transition.  After this, increasing $\dot{m}$ simply increases 
the disc (and boundary layer) temperature. The banana state is 
then analogous to the high/soft or ultrasoft state in the 
galactic black holes, but with additional luminosity from the 
boundary layer.

The disc temperature expected from a neutron star can be calculated
relative to that of a black hole emitting at the same $L/L_{Edd}$.
Given the additional boundary layer luminosity, the accretion rate
going through the black hole disc must be a factor $f$ = 2--3
higher than in the neutron star disc. 
Since the maximum disc temperature scales as
$(\dot{m}/M)^{1/4}$ and $L/L_{Edd}\propto \dot{m}$, at a given
fraction of Eddington luminosity the ratio of neutron star to black
hole disc temperatures should be $f^{-1/4} (M_{\rm BH}/M_{\rm
NS})^{1/4}$. As $M_{\rm BH}/M_{\rm NS} \approx 6$, the disc
temperature around the neutron star should be a factor 1.2--1.3 higher
than that around the black hole, for the same $L/L_{Edd}$.
The rough limit of 0.6--1.6~keV for the black hole high state disc 
temperatures translates to $\sim$ 0.8--1.9 keV for neutron stars.

Fig.~\ref{fig:ns_simple}$a$ shows a range of models which have a
Comptonized boundary layer (modelled using {\sc comptt}) with $\tau$ =
3--5 and $kT_e$ = 3 keV with seed photons from a disc at $kT_{\rm
disc}$ = 0.8--1.4 keV, with the same luminosity as the boundary
layer. This gives the same range of colours as seen on the atoll
banana branch, showing it can be explained by the increasing mass
accretion rate leading to higher disc and seed photon temperatures.

\begin{figure*}
\begin{center}
\leavevmode 
%{\psfig{file=eqpair_bh.eps,width=15cm}}
\epsfxsize=15cm \epsfbox{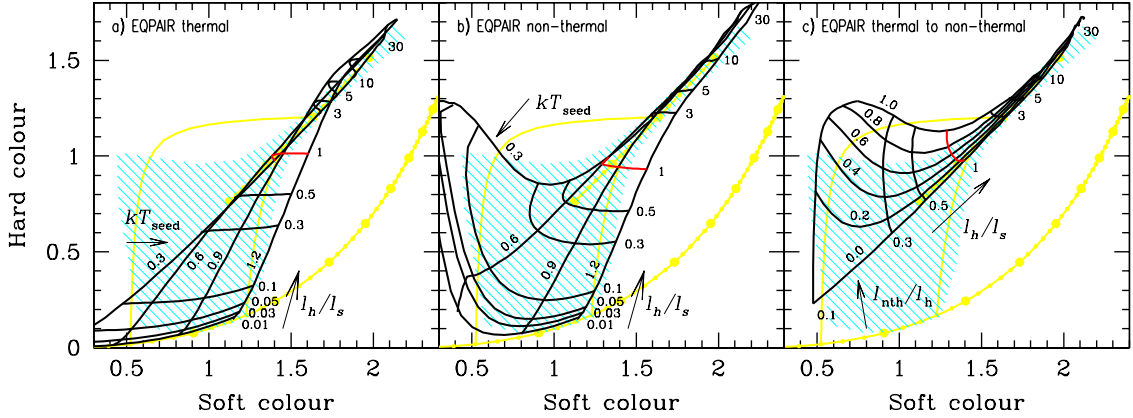}
\end{center}

\caption{Physical models for the black hole spectral evolution 
on the colour-colour diagram. The diagonally shaded area shows 
the range of colours covered by the black holes (see 
Fig.~\ref{fig:colum_bh}), while the grid shows a range of 
spectral models calculated by the {\sc eqpair} code (Coppi 1999) 
for decreasing ratio of hard (electron) luminosity, $\ell_h$ to 
soft (seed photon) luminosity, $\ell_s$ and increasing seed 
photon temperature, $kT_{\rm seed}$. Panel (a) shows the grid 
from models where the electrons are purely thermal, which cannot 
match some of the high-state spectra (e.g.\ spectrum (a) in 
Fig.~\ref{fig:colspec_bh}). Panel (b) shows that splitting the 
electron luminosity between a thermal, $\ell_{\rm th}$, and 
non-thermal, $\ell_{\rm nth}$, electron distribution with 
$\ell_{\rm nth}/\ell_{\rm h}$ fixed at 0.5 gives much better 
coverage of the observed colours. Panel (c) shows that the 
fraction of non-thermal power is not necessarily constant 
between the hard and soft states. For a fixed seed photon 
temperature of 0.3~keV the hard spectra can be explained by any 
$\ell_{\rm nth}/\ell_{\rm h}$ as Coulomb collisions efficiently 
thermalize any power-law electron injection.}

\label{fig:eqpair_bh}
\end{figure*}

\subsection{Z sources}

The Z sources are probably similar to the high luminosity 
atolls, but with the addition of a magnetic field (Hasinger \& 
van der Klis 1989). The disc evaporation efficiency decreases as 
a function of increasing mass accretion rate (e.g.\ 
R{\'o}{\.z}a{\'n}ska \& Czerny 2000), so this cannot truncate 
the disc in the Z sources. Instead the truncation is likely to 
be caused by stronger magnetic field, but here the increased 
mass accretion rate means that the inner flow/boundary layer is 
already optically thick, and so always thermalizes (Popham \& 
Sunyaev 2001; Gierli{\'n}ski \& Done 2002b).

Fig.~\ref{fig:ns_simple}$b$ shows the region of colour-colour 
space occupied by the Z sources, together with the {\sc comptt} 
banana state models from the atoll systems.  Since the 
luminosities are always close to Eddington (at least twice as 
bright as the highest luminosity seen from the atolls) we expect 
disc temperatures to be always larger then 1.4~keV.  However, a 
substantial fraction of the Z source colours lie to the {\em 
left} of the line expected for a disc temperature of 1.4~keV. 
Direct spectral fitting (e.g. Done et al.~2002) confirms that 
the disc temperature can be much lower than this, at $\sim$ 
0.8~keV for the {\it GINGA} data from Cyg~X--2. This shows that 
the low temperatures are not merely an artifact of our model 
assumptions (e.g. seed photon temperature, luminosity ratio of 
the boundary layer and disc, electron temperature) as the 
spectral fitting allowed all these to be free parameters. Much 
higher temperatures, comparable with those expected, {\em are} 
sometimes seen from Cyg~X--2 (e.g.\ di Salvo et al.~2002), from 
spectra which have large soft colour, but many of the spectra 
from both Cyg~X--2 and Sco~X--1 imply lower disc temperatures 
which are surprisingly similar to those of the lower luminosity 
atolls. While a lower disc temperature might be expected for the 
upper branch of the Z (horizontal branch) where the disc is held 
away from the neutron star by the magnetic field, this is not 
thought to be the case for the lower (flaring) branch.  One 
explanation for this could be that the inner disc is always 
slowed down from Keplarian rotation by its interaction with the 
magnetic field (e.g.\ Miller \& Stone 1997), so that its 
temperature is reduced.

%#############################################################

\section{Spectral models for the source evolution}

We have described the low/hard state source behaviour in terms 
of a truncated disc/inner hot flow, and shown that this can 
qualitatively (and to some extent quantitatively) explain the 
different spectral evolution of the neutron stars and black 
holes. However, there are alternative geometries which can 
explain the low/hard state 2--20 keV X-ray spectra, including an 
untruncated disc, with magnetic flares expanding away from the 
disc at relativistic speeds (Beloborodov 1999), or 
non-outflowing flares which strongly ionized the disc surface 
(Nayakshin, Kazanas \& Kallman 2000) or the X-rays being 
produced by synchrotron emission in a jet (e.g. Markoff et 
al.~2003). All these alternative models have severe difficulties in 
explaining the observed QPO variability without a varying inner 
disc radius, but this could merely reflect our incomplete
understanding of QPO mechanisms. 

Irrespective of the envisaged geometry, there is a clear {\em
observational} difference in spectral evolution between the black
holes and neutron stars. While the phenomenological spectral models
used in the previous section form a useful backdrop to understanding
spectral states, it is plain that more sophisticated models are
required to describe the spectra from the accretion flow. We use the
thermal/non-thermal Comptonization model of Coppi (1999), the {\sc
eqpair} code, as the basis for a more geometry independent approach to
modelling the emitted spectrum. This code has successfully fit black
hole spectra in both high/soft (Gierli{\'n}ski et al.~1999), very high
(Zdziarski et al.~2001; Gierli{\'n}ski \& Done 2002c) and low/hard
states (Zdziarski et al.~2002). It calculates the emission spectrum
resulting from Comptonization, Coulomb collisions and pair production
processes, so it can describe the spectrum formed from
these radiation processes in {\em any} geometry.

The {\sc eqpair} code assumes only that electrons with optical 
depth $\tau_p$ are accelerated by some mechanism with total 
power $\ell_h$ (parameterized as a compactness so $\ell = 
L\sigma_T / R m_ec^3$ where $L$ and $R$ are the luminosity and 
size of the region). This is split between a thermal and 
non-thermal distribution, with power $\ell_{\rm th}$ and 
$\ell_{\rm nth}$, respectively.  These electrons can cool via 
Compton scattering of soft photons (with input power $\ell_{s}$ 
at temperature $kT_s$), or via Coulomb collisions. The key 
quality of this code is that it balances these heating and 
cooling processes to calculate the {\em self-consistent} 
electron distribution and photon spectrum. Thus there is no 
specified temperature for the thermal part of the electron 
heating, because this is completely determined by the rest of 
the input parameters.  The non-thermal electrons are injected as 
a power-law with index $\Gamma_{\rm inj}$, but in general the 
low energy non-thermal electrons will cool and {\em thermalize} 
via Coulomb collisions, so even if the electron acceleration is 
predominantly non-thermal, Coulomb collisions can result in a 
predominantly thermal electron distribution which Compton 
scatters the seed photons. The code also includes the 
self-consistent pair production, so the optical depth can be 
increased when the intensity of high energy ($\ge 511$ keV) 
photons in the source is large. However, this also leads to a 
strong annihilation feature which is not seen (Grove et 
al.~1998), giving a limit on the photon density (e.g. 
Gierli{\'n}ski et al.~1999; Zdziarski et al.~2001).

\subsection{Black holes}

Fig.~\ref{fig:eqpair_bh}$a$ shows a grid of model colours from 
{\sc eqpair} assuming that the electron acceleration is 
completely thermal, for disc seed photon temperatures between 
0.3--1.2 keV (none of the black holes here are at Eddington), 
for an optical depth of unity and $\ell_s=1$.  In the low/hard 
state the disc temperature is observed to be generally lower 
than $0.6$ keV, and the observed hard spectrum requires 
$\ell_h/\ell_s\gg1$.  Both these are naturally produced in a 
truncated disc model (disc is far away, so has low temperature, 
and few disc photons illuminate the central source), but the 
same resulting spectrum is produced for any geometry which has 
the same seed photon temperature and $\ell_h/\ell_s$. There is a 
limit to how few seed photons can be present, which we take to 
be $\ell_h/\ell_s=30$, as cyclo-synchrotron photons produced in 
the flow can become the dominant source of seed photons (e.g. 
Wardzi{\'n}ski \& Zdziarski 2000). Conversely, to produce the 
high/soft state spectra requires much smaller $\ell_h/\ell_s$, 
so we show the colours for a grid of models with $\ell_h/\ell_s$ 
from 0.01 to 30.

Plainly the thermal models with temperatures below 0.6~keV and
$\ell_h/ \ell_s > 1$ reproduce the low/hard state colours well.  The
self consistently derived electron temperature from {\sc eqpair} is
around 100 keV, so is compatible with the higher energy data as well.
Changing the disc temperature below 0.6~keV makes little difference to
the low/hard state colours as the spectrum is effectively a power-law.
However, for $\ell_h/\ell_s < 1$ the range of observed soft state
colours is not well reproduced by these thermal models. Regions in the
colour-colour diagram corresponding to the classic high state spectra
i.e. with soft colour around 0.5 and hard colour around 1 are
inaccessible with a simple thermal plasma. Comparison with the spectra
shown in Fig.~\ref{fig:colspec_bh} shows that the problem is that the
disc spectrum is much more dominant in the data than in our models. We
assumed $\tau=1$, so less than half of the seed photons can escape
without scattering, and the scattered spectrum is always of comparable
luminosity to that of the disc. The relative importance of the disc
emission can be easily increased by either reducing the optical depth,
or by considering a geometry in which many of the seed photons from
the disc are not intercepted by the Comptonizing region (e.g. if the
hot electrons are in magnetic flares which cover only a small fraction
of the disc surface). However, while this would match the colours used
here, detailed spectral fitting which includes data at higher energies
clearly shows the presence of a non-thermal tail (e.g. Gierli{\'n}ski
et al.~1999; Zdziarski et al.~2001; Gierli{\'n}ski \& Done 2002c),
while the thermal models with $\ell_s/\ell_h > 0.5$ have $kT_e\le 30$
keV. Guided by these results, Fig.~\ref{fig:eqpair_bh}$b$ shows the
same set of models but using a hybrid thermal/non-thermal electron
distribution with $\ell_{nth}/\ell_h$ = 0.5. Clearly this can
reproduce the whole range of colours seen in the black hole binaries.

The ratio of non-thermal to thermal power is not necessarily 
constant from the hard to soft states. 
Fig.~\ref{fig:eqpair_bh}$c$ shows a grid of models with 
$\ell_{nth}/\ell_h$ ranging from 0 to 1 for a fixed disc 
temperature of 0.3 keV.  Plainly the hard state is completely 
insensitive to the non-thermal fraction (Maccarone \& Coppi 
2002a).  This is because the electron distribution is calculated 
self-consistently in the {\sc eqpair} code by balancing heating 
(non-thermal injection) and cooling (Compton scattering and 
Coulomb collisions).  If Coulomb collisions dominate the cooling 
then the electrons thermalize, irrespective of their initial 
injected distribution. In the low state Coulomb scattering 
dominates the cooling as Compton scattering is limited due to 
the lack of seed photons ($\ell_s < \ell_h$). Conversely, in the 
high state, Compton scattering dominates as seed photons are 
plentiful, so the electron distribution can retain its 
non-thermal character. To describe the spectra with soft and 
hard colours of 0.5 and 1 (that are {\em never} observed in 
neutron star systems) {\em requires} a low temperature disc and 
a considerable fraction of power injected as non-thermal 
electrons.

\subsection{Atolls}

\begin{figure}
\begin{center}
\leavevmode
%{\psfig{file=transform.eps,width=6cm}}
\epsfxsize=6cm \epsfbox{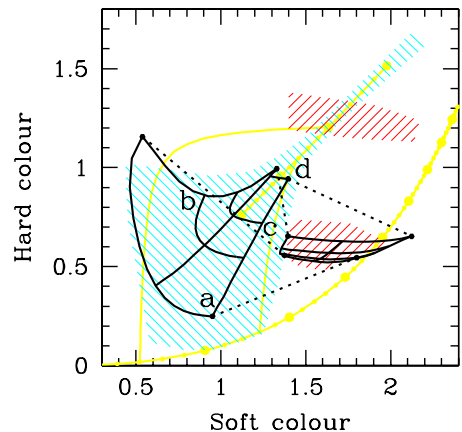}
\end{center}

\caption{Transformation of colours of the hybrid plasma black 
hole models (Fig.~\ref{fig:eqpair_bh}$b$) due to additional 
emission from the boundary layer. The grid on the left is a 
fragment of the grid in Fig.~\ref{fig:eqpair_bh}, restricted in 
$kT_{\rm seed}$ and $\ell_h/\ell_s$. The grid on the right was 
created using the same models, but with an additional emission 
from cold, optically thick plasma ($kT_e$ = 2.2 keV, $\tau$ = 6 
with $kT_{\rm seed}=1.5 kT_{\rm disc}$), representing the 
contribution from the boundary layer. The luminosity of the 
boundary layer is set to be equal to the luminosity of the main 
component. The area in the colour-colour diagram covered by the 
soft states of black hole binaries (left grid) is shifted to the 
area typical for banana states of atolls due to the emission 
from the boundary layer. The very presence of the boundary layer 
and lack of the horizon in neutron stars prevents them from 
entering the area covered by the grid on the left. }

\label{fig:transform}
\end{figure}

Fig.~\ref{fig:transform} shows the resulting colours assuming 
that the accretion flow in atolls is {\em identical} to that in 
black holes but with the addition of an optically thick boundary 
layer, which we model using the {\sc comptt} code. This is only 
applicable at high mass accretion rates (as otherwise the 
boundary layer is optically thin), so we include only the black 
hole soft state spectral models. We use the hybrid plasma 
(Fig.~\ref{fig:eqpair_bh}$b$) with seed photons temperature of 
0.3--0.9 keV and $\ell_h/\ell_s$ between 0.1 and 1 to model the 
accretion flow. For the boundary layer we use thermal 
Comptonization in plasma with $\tau$ = 6 and $kT_e$ = 2.2 keV. 
The seed photons for the boundary layer (assumed to origin from 
the neutron star surface) have temperature $1.5$ times that of 
the seed photons for the hybrid plasma (from the disc). The wide 
region of soft colours covered by the black hole models is 
transformed into a narrow strip around the atoll banana branch 
simply by the addition of the optically thick boundary layer. 

\begin{figure*}
\begin{center}
\leavevmode
%{\psfig{file=transpec.eps,width=16cm}}
\epsfxsize=16cm \epsfbox{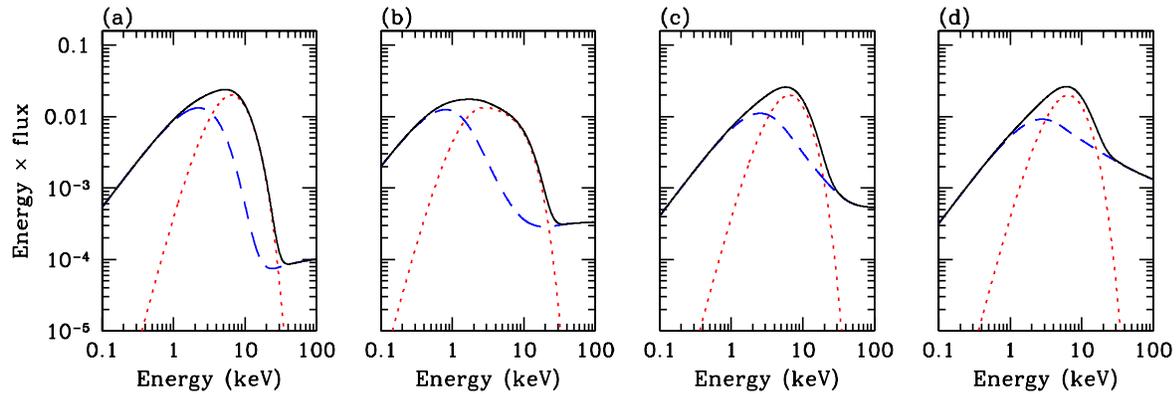}
\end{center}

\caption{{\sc eqpair} model spectra from the points marked in 
Fig.~\ref{fig:transform} (dashed curve) with additional emission from 
the boundary layer (dotted curve). The total spectra correspond to 
the shifted points (right-hand grid) in Fig.~\ref{fig:transform}. The 
{\sc eqpair} model parameters are: (a) $kT_{\rm disc}$ = 0.9 keV, 
$\ell_s/\ell_h$ = 0.1, (b) $kT_{\rm disc}$ = 0.3 keV, $\ell_s/\ell_h$ 
= 0.3, (c) $kT_{\rm disc}$ = 0.9 keV, $\ell_s/\ell_h$ = 0.5, (d) 
$kT_{\rm disc}$ = 0.9 keV, $\ell_s/\ell_h$ = 1. Only the latter 
spectrum has a hard tail which is incompatible with the level of 
detections and upper limits for high energy neutron star emission at 
high mass accretion rates.}

\label{fig:transpec}
\end{figure*}

Fig.~\ref{fig:transform} shows the resulting spectra from four 
of these models (points $a$, $b$, $c$ and $d$ in 
Fig.~\ref{fig:transform}), with the hybrid plasma from the 
accretion flow as the blue line and the boundary layer as a red 
line.  Plainly these all predict that hard tails should be seen 
in neutron stars systems, and indeed recent {\it BeppoSAX\/} and 
{\it RXTE\/} data have detected tails at comparable levels to 
that shown in the ultrasoft, high state and moderate very high 
state spectra (panels $a$, $b$ and $c$ in 
Fig.~\ref{fig:transpec}), but the extreme very high state 
spectra (Fig.~\ref{fig:transpec}$d$) seems stronger than any yet 
seen in the neutron star systems (e.g. di Salvo et al.~2001 and 
references therein). This is plausibly due to the increased 
Compton cooling of the accretion flow from the boundary layer 
emission, giving neutron stars a lower $\ell_h/\ell_s$ ratio 
than black holes, which leads to a steeper/weaker hard tail.

\section{Conclusions}

The huge amount of data taken from {\it RXTE\/} makes it possible to
study the average spectral evolution of black holes versus that from
neutron stars. With intrinsic colours, data from different objects
can be directly compared, showing that black hole spectra clearly
evolve very differently from those of neutron stars with increasing
average flux. An unknown transient system can be classified as a black
hole or neutron star simply in terms of its colour evolution as a
function of flux.

We show that this spectral evolution is consistent with models of the
accretion flow which involve a truncated disc at low mass accretion
rates, where the inner disc is replaced by an X-ray hot flow. However,
irrespective of the underlying geometry, we show that we can match the
evolution of the broad band spectral shape in the accreting black
holes in terms of thermal/non-thermal Comptonization models for the
X-ray emission, and that these {\em same} emission models can also
match the very different neutron star spectral evolution.  The key to
their different behaviour is simply that neutron stars have a solid
surface, giving an additional emission component from the boundary
layer, while black holes do not.

\section*{Acknowledgements}

We thank Didier Barret and Aya Kubota for useful discussions.

%-----------------------------------------------------

\label{lastpage}


\begin{thebibliography}{}

%\bibitem[]{arn96} Arnaud K. A., 1996, in Jacoby G. H., Barnes J., eds.,
%Astronomical Data Analysis Software and Systems V. ASP Conf. Series Vol.\ 101,
%San Francisco, p.\ 17

%\bibitem[]{che99} Chevalier C., Ilovaisky S. A., Leisy P., Patat F., 
%1999, A\&A, 
%347, L51


\bibitem[]{ba82} Backer D. C., Kulkarni S. R., Heiles C.,
Davis M. M., Goss W. M., 1982, Nature, 300, 61

\bibitem[]{b95} Ba{\l}uci{\'n}ska-Church M.,Belloni T., Church M. J.,
Hasinger G., 1995, A\&A, 309, L5 

\bibitem[]{b96} Barret D., McClintock J. E., Grindlay J. E.,  1996,
ApJ, 473, 963

\bibitem[]{b00} Barret, D.; Olive, J. F.; Boirin, L.; Done, C.;
Skinner, G. K.; Grindlay, J. E. 2000, ApJ, 533, 329

\bibitem[]{bv94} Barret D., Vedrenne G.,  1994, ApJS, 92, 505

\bibitem[]{bpv02} Belloni T., Psaltis D., van der Klis M., 
2002, ApJ, 572, 392

\bibitem[]{b99} Beloborodov A.M. 1999, ApJ, 510, L123

\bibitem[]{bfg99} Bradshaw C. F., Fomalont E. B.,  Geldzahler B. J.,
1999, ApJ, 512, L121 

\bibitem[Campana et al.~2002]{2002ApJ...575L..15C} Campana S., Stella L., 
Gastaldello F., et al., 2002, ApJ,  575, L15 

\bibitem[]{cgr00} Churazov E., Gilfanov M., Revnivtsev M., 2001, 
MNRAS, 321, 759

\bibitem[]{cbc01} Church M. J., Ba{\l}uci{\'n}ska-Church M., 2001,
A\&A, 369, 915

\bibitem[Cowley et al.~1983]{1983ApJ...272..118C} Cowley A.~P., 
Crampton D., Hutchings J.~B., Remillard R., Penfold J.~E., 1983, 
ApJ,  272, 118 

\bibitem[]{cow02} Cowley A. P., Schmidtke P. C., Hutchings J. B.,
Crampton D., 2002, AJ, 123, 1741

\bibitem[]{cop99} Coppi P. S., 1999, in ASP Conf. Ser. 161, High 
Energy Processes in Accreting Black Holes, ed. J. Poutanen \& R. 
Svensson (San Francisco: ASP), 375

%\bibitem[]{cui98} Cui W., Barret D., Zhang S. N., Chen W., Boirin 
%L.,
%Swank J., 1998, ApJ, 502, L49

\bibitem[]{czc98} Cui W., Zhang S. N., Chen W.,  1998, ApJ, 492, L53

\bibitem[]{dib97} Di Benedetto G. P. 1997, ApJ, 486, 60

\bibitem[]{dis00} Di Salvo T., Iaria R., Burderi L., Robba N. R.,
2000a, ApJ, 542, 1034

\bibitem[Di Salvo et al.~2000]{2000ApJ...544L.119D} Di Salvo T., 
Stella L., Robba N.~R., et al., 2000b, ApJ,  544, L119 


%\bibitem[]{dis01a} Di Salvo T., M{\'e}ndez M., van der Klis M., Ford 
%E., Robba N. R., 2001a, ApJ, 546, 1107

\bibitem[]{dis01b} Di Salvo T., Robba N. R.,Iaria R., Stella L., 
Burderi L., Israel G. L., 2001, 554, 49

\bibitem[]{dis02} Di Salvo T., et al., 2002, A\&A, 386, 535

\bibitem{d92}  Done C., Mulchaey J.S., Mushotzky R.F., Arnaud K.A., 1992,
ApJ, 395, 275

\bibitem[Done et al.~2002]{2002MNRAS.331..453D} Done C., {\. 
Z}ycki P.~T., Smith D.~A., 2002, MNRAS,  331, 453 

\bibitem[]{dhl01} Dubus G., Hameury J.-M., Lasota J.-P., 2001, A\& A,
373, 251

\bibitem[]{emn97} Esin A.A., McClintock J.E., Narayan R. 1997, ApJ, 489, 865

%\bibitem[]{fk01} Fender R. P., Kuulkers E., 2001, MNRAS, 324, 923

%\bibitem[]{fkk98} Ford E. C., van der Klis M., Kaaret P., 1998, ApJ, 
%498, L41

%\bibitem[]{for99}  Ford E. C., van der Klis M., M{\'e}ndez M., van 
%Paradijs J., 
%Kaaret P., 1999, ApJ, 512, L31 

%\bibitem[]{flm89} Fortner B., Lamb F. K., Miller G. S., 1989, 
%Nature, 342, 775

\bibitem[]{g01} Garcia M. R., McClintock J. E., Narayan R., Callanan
P., Barret D., Murray S. S., 2001, ApJ, 553, L47

\bibitem[]{gme01} Gierli{\' n}ski M., 
Macio{\l}ek-Nied{\' z}wiecki A., Ebisawa K., 2001, MNRAS, 325, 1253

\bibitem[]{gd02a} Gierli{\'n}ski M., Done C., 2002a, MNRAS, 331, L47

\bibitem[]{gd02b} Gierli{\'n}ski M., Done C., 2002b, MNRAS, 337, 
1373

\bibitem[]{gd02c} Gierli{\'n}ski M., Done C., 2002c, MNRAS, 
submitted (astro-ph/0212384)

\bibitem[]{gie99} Gierli{\'n}ski M., Zdziarski A. A., Poutanen J., 
Coppi P. S., Ebisawa K., Johnson W. M., 1999, MNRAS, 309, 496

\bibitem[]{2001MNRAS.325.1253G} Gierli{\'n}ski M., 
Macio{\l}ek-Nied{\'z}wiecki A., Ebisawa K., 2001, MNRAS,  325, 
1253 

\bibitem[]{gcr99} Gilfanov M., Churazov E., Revnivtsev M., 1999, A\&
A, 352, 182

\bibitem[]{gm77} Goss W. M., Mebold U., 1977, MNRAS, 181, 255

\bibitem[Grove et al.~1998]{1998ApJ...500..899G} Grove J.~E., 
Johnson W.~N., Kroeger R.~A., McNaron-Brown K., Skibo J.~G., 
Phlips B.~F., 1998, ApJ,  500, 899 


\bibitem[]{haar01} Haardt F., et al.~2001, 2001, ApJS., 133, 187

\bibitem[]{ht94} Haberl F., Titarchuk L., 1994, A\&A, 299, 414

\bibitem[]{hk89} Hasinger G., van der Klis M., 1989, A\&A, 225, 79

\bibitem[]{hb02} Hawley J. F., Balbus S. A., 2002, ApJ, 573, 738

\bibitem[]{h95} Herrero A., Kudritzki R. P., Gabler R., Vilchez
J. M., Gabler A., 1995, A\&A, 309, 556 

\bibitem[]{hr95} Hjellming R. M., Rupen M. P., 1995, Nat, 375, 464

\bibitem[]{hom01} Homan J., Wijnands R., van der Klis M., 
Belloni T., van Paradijs J., Klein-Wolt M., Fender R.,
M{\'e}ndez M., 2001, ApJS, 132, 377

\bibitem[]{hom02} Homan J., van der Klis M., Jonker P. G., Wijnands
R., Kuulkers E., M{\'e}ndez M., Lewin W. H. G., 2002, ApJ, 568, 878

%\bibitem[]{lam89} Lamb F. K., 1989, in Proc. 23 ESLAB Symp. on X-Ray
%Astronomy, ed. N. E. White, ESA SP-296 (Noordwijk: ESA)


\bibitem[]{h02} Hynes R. I., Haswell C. A., Chaty S.,
Shrader C. R.,  Cui W., 2002, MNRAS, 331, 169

\bibitem[]{zan01} in 't Zand J. J. M., et al.~2001, astro-ph/0104285

\bibitem[]{kl87} King A. R., Lasota J.--P.,  1987, A\&A, 185, 155

\bibitem[]{kr98} King A.R., Ritter H., 1998, MNRAS, 293, L42

\bibitem[]{kks97} King A. R., Kolb U., Szuszkiewicz E., 1997, ApJ, 
488, 89

\bibitem[]{k94} Kuulkers E., van der Klis M., Oosterbroek T., Asai K., 
Dotani T., van Paradijs J., Lewin W. H. G.,  1994, A\&A, 289, 795

\bibitem[]{k96} Kuulkers E., van der Klis M., Vaughan B. A., 1996,
A\&A, 311, 197

\bibitem[]{k98} Kuulkers E., 
Wijnands R., Belloni T., Mendez M., van der Klis M., van Paradijs
J., 1998, ApJ, 494, 753 

\bibitem[]{k02} Kuulkers E., Homan J., 
van der Klis M., Lewin W.~H.~G., M{\' e}ndez M., 2002, A\&A,  382, 947 

%\bibitem[]{lan87} Langmeier A., Sztajno M., Hasinger G., Tr{\"u}mper 
%J., 
%Gottwald M., 1987, ApJ, 323, 288

\bibitem[]{lt99} Laurent P., Titarchuk L., 1999, ApJ, 511, 289

%\bibitem[]{mc02} Maccarone T. J., Coppi P., 2002a, MNRAS, in press
%(astro-ph/0209116) 

\bibitem[]{mc02a} Maccarone T. J., Coppi P., 2002a, MNRAS, 
submitted (astro-ph/0204235)

\bibitem[]{mc02b} Maccarone T. J., Coppi P., 2002b, MNRAS, 
submitted (astro-ph/0209116)

\bibitem[]{m03} Markoff S., Nowak M., Corbel S., Fender R., \& Falcke
H.\ 2003, A\&A, 397, 645

\bibitem[]{m01} McClintock J.E., et al., 2001, ApJ., 555, 477

\bibitem[]{mn01} Medvedev M. V., Narayan R.,  2001, ApJ, 554, 1255

\bibitem[]{m99} Menou K., Esin A. A., Narayan R., Garcia M. R., 
Lasota J.-P., McClintock J. E., 1999, ApJ, 520, 276

\bibitem[]{mfr00} Merloni A., Fabian A. C., Ross R. R.,  2000, 
MNRAS, 313, 193

\bibitem[]{mig02} Mignani R. P., De Luca A., Caraveo P. A., Mirabel
I. F., 2002, A\&A, 386, 487

\bibitem[]{ms97} Miller K.A., Stone J.M.,  1997, ApJ, 489, 890

\bibitem[]{mit84} Mitsuda K., et al.~1984, PASJ, 36, 741

\bibitem[]{mit89} Mitsuda K., Inoue H., Nakamura N., Tanaka Y., 1989,
PASJ, 41, 97

\bibitem[Muno et al.~2002]{2002ApJ...568L..35M} Muno M.~P., 
Remillard R.~A., Chakrabarty D., 2002, ApJ,  568, L35 

\bibitem[]{nak89} Nakamura N., Dotani T., Inoue H., Mitsuda K., Tanaka Y., 
Matsuoka M.,1989, PASJ, 41, 617

\bibitem[]{ny95} Narayan R., Yi I., 1995, ApJ., 452, 710

\bibitem[]{n97} Narayan R., Garcia M. R., McClintock J. E., 1997, ApJ,
478, 79

\bibitem[]{nkk00} Nayakshin S., Kazanas D., Kallman T.R., 2000, ApJ.,
537, 833

\bibitem[]{n95} Nowak M.A., 1995, PASP, 107, 1207

\bibitem[]{n00} Nowak M. A., 2000, MNRAS, 318, 361

\bibitem[]{ob97} Orosz, J.~A. Bailyn, C.~D., 1997, ApJ, 477, 876 

\bibitem[]{ok99} Orosz J.~A., Kuulkers E., 1999, MNRAS,  305, 132 

\bibitem[]{o02} Orosz J. A., et al., 2002, ApJ, 568, 845

\bibitem[]{pkw98} Paerels F., Kahn S. M., Wolkovitch D. N., 
1998, ApJ., 496, 473

\bibitem[]{pen89} Penninx W., Damen E., Tan J., Lewin W. H. G., van
Paradijs J., 1989, A\&A, 208, 146

\bibitem[]{ps01} Popham R., Sunyaev R., 2001, ApJ, 547, 355

\bibitem[]{ps95} Predehl P., Schmitt J. H. M. M., 1995, A\&A, 293, 889

\bibitem[]{pn01} Psaltis D., Norman C., 2000, astro-ph/0001391

\bibitem[]{rc00} R{\'o}{\.z}a{\'n}ska A., Czerny B., 2000, MNRAS, 
316, 473

\bibitem[]{rut01} Rutledge R. E., Bildsten L., Brown E. F., Pavlov G. 
G., Zavlin V. E., 2001, ApJ, 559, 1054

\bibitem[]{ss00} Sibgatullin N. R., Sunyaev R. A.,  2000, AstL, 26, 699

\bibitem[]{sle76} Shapiro S.L., Lightman A.P., Eardley D.M. 
1976, ApJ, 204, 187

\bibitem[]{st95} Shimura T., Takahara F.,  1995, ApJ, 445, 780

\bibitem[]{sbl99} Shirey R. E., Bradt H. V., Levine A. M.,  1999, ApJ,
517, 472

\bibitem[]{sv98} Stella L., Vietri M., 1998, ApJ, 492, L59

\bibitem[]{ss86} Sunyaev R. A., Shakura N. I., 1986, SvAL, 12, 117

\bibitem[]{s91} Sunyaev R. A. et al., 1991, SvAL,17, 409

\bibitem[]{sr00}  Sunyaev R. A., Revnivtsev M., 2000, A\&A, 358, 617

\bibitem[]{tl95} Tanaka Y., Lewin W. H. G.  1995, in X--Ray Binaries,
ed. W. H. G. Lewin, J. van Paradijs \& E. van den Heuvel (Cambridge:
Cambridge Univ. Press), 126

\bibitem[]{t86} Tennant A. F., Fabian A. C., Shafer R. A., 1986,
MNRAS, 221, L27

\bibitem[]{kli95} van der Klis M., 1995, in Lewin W. H.  G., van
Paradijs J., van den Heuvel E. P. J., eds., X-ray Binaries, Cambridge
University Press, Cambridge, P.\ 252

\bibitem[]{kli00} van der Klis M., 2000, ARAA, 38, 717

\bibitem[]{kli01} van der Klis M., 2001, ApJ, 561, 943

\bibitem[]{Wac02} Wachter S., Hoard D. W., Baylin C. D., Corbel S.,
Kaaret P., 2002, ApJ., 568, 901

\bibitem[]{wz01} Wardzi{\'n}ski G., Zdziarski A. A., 2000, MNRAS, 314,
183

\bibitem[]{wk99} Wijnands R., van der Klis M., 1999, ApJ, 514, 939 

\bibitem[]{wm02} Wijnands R., Miller J. M., 2002, ApJ, 564, 974

\bibitem[]{wd01} Wilson C. D., Done C., 2001, MNRAS, 325, 167

\bibitem[]{whi88} White N. E., Stella L., Parmar A. N., 1988, ApJ, 324, 363

\bibitem[]{wm84} White N. E., Marshall F. E., 1984, ApJ, 281, 354

\bibitem[]{whi84} White N. E., Parmar A. N., Sztajno M., Zimmermann
H. U., Mason K. O., Khan S. M., 1984, ApJ, 283, L9

\bibitem[]{Yos93} Yoshida K., Mitsuda K., Ebisawa K., Ueda Y., Fujimoto R., 
Taqoob T., Done C., 1993, PASJ, 45, 605

\bibitem[]{Z98} Zdziarski A. A., Poutanen J., Mikolajewska J.,
Gierli{\'n}ski M.,  Ebisawa K., Johnson W. N., 1998, MNRAS, 301, 435 

\bibitem[]{zdz01} Zdziarski A. A., Grove J. E., Poutanen J., Rao A. 
R., Vadawale S. V., 2001, ApJ, 554, L45

\bibitem[]{zdz02} Zdziarski A. A., Poutanen J., Paciesas W. S., Wen 
L.,  2002, ApJ, 578, 357

\bibitem[]{zcc97} Zhang S. N., Cui W., Chen W., 1997, ApJ, 482, L155

\bibitem[]{zds98} {\.Z}ycki P.T., Done C., Smith D.A., 1998, 
ApJ, 496, L25

\end{thebibliography}
\end{document}